\documentclass[aps,superscriptaddress,showpacs,nofootinbib]{revtex4}
\usepackage{graphics}
\usepackage{bm}

\input{epsf}

\newcommand{\be}{\begin{equation}}
\newcommand{\ee}{\end{equation}}
\newcommand{\bea}{\begin{eqnarray}}
\newcommand{\eea}{\end{eqnarray}}

\begin{document}

\preprint{UG-FT-245/09}

\preprint{CAFPE-115/09}

\title{Warm inflation model building}
\author{Mar Bastero-Gil}
\email{mbg@ugr.es}
\affiliation{Departamento de F\'{\i}sica Te\'orica y del Cosmos,
  Universidad de Granada, Granada-18071, Spain}

\author{Arjun Berera}
\email{ab@ph.ed.ac.uk}
\affiliation{ School of Physics and Astronomy, University of
Edinburgh, Edinburgh, EH9 3JZ, United Kingdom}

\begin{abstract}
We review the main aspects of the warm inflation scenario, focusing
on the inflationary dynamics and the predictions related to the
primordial spectrum of perturbations, to be compared with the recent
cosmological observations. We study in detail three different classes of
inflationary models, chaotic, hybrid models and hilltop models, and
discuss their embedding into supersymmetric models and the
consequences for model building of the warm inflationary dynamics
based on first principles calculations. Due
to the extra friction term introduced in the inflaton background
evolution generated by the dissipative dynamics, inflation can take
place generically for smaller values of the field, and larger values
of couplings and masses. When the dissipative
dynamics dominates over the expansion, in the so-called strong
dissipative regime, inflation proceeds with sub-planckian inflaton
values. Models can be naturally embedded into a supergravity framework, with
sugra corrections suppressed by the Planck mass now under control, for
a larger class of K\"ahler potentials. In particular, this provides a
simpler solution to the ``eta'' problem in supersymmetric hybrid 
inflation, without restricting the K\"ahler potentials compatible with
inflation.   For chaotic models dissipation leads
to a smaller prediction for the tensor-to-scalar ratio and a less
tilted spectrum when compared to the cold inflation scenario. We find
in particular that a small component of dissipation renders the
quartic model now consistent with the current CMB data.

\medskip                                   

\noindent
keywords: cosmology, inflation, dissipation, fluctuation, supersymmetry.
\end{abstract}
                                                                                \pacs{98.80.Cq, 11.30.Pb, 12.60.Jv}

\maketitle

\medskip

Invited review for International Journal of Modern Physics A

\section{Introduction}
\label{intro}

Warm inflation achieves a few milestones in the development
of inflationary cosmology.  
First it is the only example of an inflationary dynamics in
which the state of the universe during inflation is
not the vacuum state, as in the standard picture, but rather
an excited statistical state, with the thermal state being
the one most examined.
Second there are several first principles
warm inflation models that by now exist, which are consistent
both with cosmological observation and quantum field theory.
For the standard inflation dynamics, when accounting for all
aspects of the scenario up to the end of the reheating phase,
there have been only a limited number of first principles models
developed.  
Third only in warm inflation dynamics are the
simplest monomial potential models consistent.
Fourth, warm inflation provides an alternative solution
to the graceful exit problem to the standard inflation dynamics.
Finally warm inflation models are free of the "$\eta$"-problem,
which has perplexed model building in standard inflation
dynamics.

Inflation remains today the most attractive solution to
the cosmological puzzles, in particular the horizon and
flatness problems.  To date there have been two distinct
dynamical realizations of inflation.
In the original picture of inflation 
\cite{Guth:1980zm,Sato:ds,Albrecht:1982wi,Linde:1981mu,Linde:1983gd},
which has become the standard inflation picture,
the universe rapidly supercools during inflation and
subsequently a reheating phase is invoked to end inflation and
put the universe back into a radiation dominated regime.
This picture is termed
cold, supercooled or isentropic
inflation.
In the other picture, termed warm or nonisentropic inflation
\cite{Berera:1995ie},
dissipative effects are important during the inflation period,
so that radiation production occurs concurrently with inflationary
expansion.

The idea of inflationary expansion and particle production occurring
concurrently was first suggested by L.Z. Fang in his pre-inflation
inflation paper in 1980 \cite{Fang:1980wi}. Later in the 80s,
the suggestion was made by Moss \cite{im} and
Yokoyama and Maeda \cite{Yokoyama:1987an} to
introduce an $\Upsilon {\dot \phi}$ term in the inflaton
evolution equation, as a source of radiation production.
This idea was then independently rediscovered almost a decade later 
by Berera and Fang \cite{Berera:1995wh}
who went further to suggest that the dynamics
of the inflaton field be governed by a Langevin equation
which not only included a $\Upsilon {\dot \phi}$ dissipative
term but also a noise force term that would drive
the inflaton fluctuations, with a fluctuation-dissipation
theorem uniquely specifying the inflaton fluctuations. 
Their paper established the
foundations for the theory of fluctuations in warm inflation.
Subsequently \cite{Berera:1995ie} proposed
completely eliminating the reheating phase, which
up to then was present in
all inflation models.  The new scenario
was that the process of radiation production
during inflation could occur under strong enough dissipation
to permit a slow-roll solution within the inflationary
expansion period
and subsequently enough radiation could be present
to end the inflation phase and smoothly enter the radiation
dominated phase.
This presented an alternative solution to the graceful
exit problem.
To complete the development of the basic warm inflation scenario,
the calculations of inflaton fluctuations for an inflaton
governed by a Langevin equation was done in \cite{Berera:1999ws}.

The interesting features of the warm inflation scenario have
motivated understanding the first principles origin of
such a dynamics.  The quantum mechanical origin of
the Langevin equation for warm inflation, and the
fluctuation-dissipation relation it implies were studied
in \cite{Berera:1996nv}.  The quantum field theory origin of the strong
dissipation needed to realize warm inflation was first
examined by Berera, Gleiser and Ramos \cite{BGR}.
This work proposed that solutions relevant to
warm inflation should be explored within an adiabatic
approximation within quantum field theory, which is
the approximation that has been followed by all subsequent work
in this area.
Berera and Ramos \cite{BR1} then proposed a two-stage interaction
configuration which could yield strong dissipation
as well as permit the flat potentials needed
for inflation.  In this model the inflaton field was
directly coupled to very heavy Bose and Fermi
``catalyst'' fields, much heavier than the temperature
scale in the universe, and these heavy fields were in turn coupled
to light fields.   The heaviness of the catalyst
fields meant they were essentially in a zero-temperature
state, and so in SUSY models the primary quantum loop
corrections would cancel, thereby maintaining the
flat potentials needed for inflation.  Nevertheless
the heavy fields acted to catalyze the production of
light fields which thereby heated the universe.
This two-stage mechanism alleviated concerns in realizing warm inflation
from quantum field theory raised in \cite{BGR}
and then by Yokoyama and Linde \cite{Yokoyama:1998ju}.
The first calculation of
dissipative coefficients for the two-stage mechanism
were done by Moss and Xiong \cite{mx}.

The development of a quantum field theory basis for warm inflation
led to specific warm inflation models.  The first model
that was both consistent with cosmological
constraints and quantum field theory was proposed
for a case where all the fields were in a 
high temperature phase \cite{BGR2}, but
the model required a large number of fields and
a non-standard configuration of interactions.
Much more successes was obtained in implementing
models with the two-stage mechanism.
The first attempt at building first principles
warm inflation models for common potentials
such as monomial and hybrid, was done by
Bastero-Gil and Berera \cite{bb4}.
Several such models have been developed and in
this review they will be presented in detail.

There have been two earlier reviews of warm inflation.
In \cite{Berera:2006xq} the basics of the scenario
were covered and in \cite{Berera:2008ar} the quantum
field theory dynamics was covered.  In this review,
the various first principles models of warm inflation
that have been developed will be examined and further new
analysis of these models will be done.
This review is organized as follows.  In
Sect. \ref{sect2} the basic warm inflation scenario is
presented.  In Sects. \ref{wimonomial} - \ref{wihilltop}
first principles models of warm inflation for monomial,
hybrid, and hilltop potentials respectively are examined.
In Sect. \ref{wimodelb}, particle physics model building
consequences of warm inflation in these models is discussed.
Finally, in
Sect. \ref{summary} we summarize the main results
and discuss future directions of development.

\section{Warm inflation: background field evolution}
\label{sect2}

In the standard cold inflation picture, the inflaton field is taken as
an isolated system, such that its potential energy density drives the quasi
exponential expansion of the universe, independently of any other form
of energy density which could be initially present. Indeed, any other
initial component of the energy density, like radiation, will be
quickly redshifted away during inflation. Under this assumption,
calling $\phi(t)$ the background value of the inflaton field, with
potential energy density $V(\phi)$, the inflationary evolution is
given by the equation:
\be
\ddot \phi(t) + 3 H \dot \phi(t) + V_\phi \simeq 0 \,,  
\ee
with the expansion rate\footnote{Throughout this article we will use
  the reduce Planck mass $m_P \simeq 2.4 \times 10^{18}$ GeV.} $H^2= (
\dot \phi^2/2 + V(\phi))/(3 m_P^2)$, and 
$V_\phi$ being the derivative of the potential energy with respect to
the field. Inflation, i.e. accelerated expansion, happens when $\dot
\phi^2 \ll V(\phi)$, such that $3 H^2 m_P^2 \simeq V(\phi)$ and $\ddot
\phi(t) \ll H \dot \phi(t)$, and the equation of  motion of the inflaton
field reduces to:  
\be
3 H \dot \phi + V_\phi \simeq  0 \,,   
\ee
i.e., its motion is overdamped by the expansion. 
The consistency conditions for the approximations to hold are given by
the so-called slow-roll conditions:
\bea
\epsilon_\phi &=& \frac{m_P^2}{2} \left ( \frac{V_{\phi}} {V}
\right)^2 \ll 1\,, \label{epsphi}\\
\eta_\phi &=& m_P^2 \left ( \frac{V_{\phi \phi}} {V}\right) \ll 1 \,. \label{etaphi}
\eea
Nevertheless in any particle
physics realisation of the inflationary framework, the inflaton is not
an isolated part of the model but it interacts with other fields. The
cold inflation scenario assumes that those interactions have no effect 
on the inflationary dynamics and can be neglected during
inflation. However, they may 
lead to the dissipation of the inflaton energy into other degrees of
freedom, such that a small percent of the inflaton vacuum energy is
transferred into other, although subdominant, kinds of energy. This is
the essence of the warm inflaton scenario
\cite{Berera:1995ie,Berera:1996nv,Berera:1996fm}. 
The dissipative term appears as an
extra friction term in the evolution equation for the inflaton field $\phi$,
\be
\ddot \phi + ( 3 H + \Upsilon ) \dot \phi + V_\phi =0
\,,\label{eominf}
\ee
with $\Upsilon$ being the dissipative coefficient. Eq. (\ref{eominf})
is equivalent to the evolution equation for the inflaton energy
density $\rho_\phi$:
\be
\dot \rho_\phi + 3 H ( \rho_\phi + p_\phi) = - \Upsilon ( \rho_\phi +
p_\phi) \,,
\label{rhoinf}
\ee
with pressure $p_\phi = \dot \phi^2/2 - V(\phi)$, and $\rho_\phi
+ p_\phi= \dot \phi^2$. Energy conservation then demands that the
energy lost of the inflaton field must be gained by some other fluid
component $\rho_\alpha$, with the  RHS of Eq. (\ref{rhoinf}) acting as
the source term:
\be
\dot \rho_\alpha+ 3 H ( \rho_\alpha + p_\alpha) = \Upsilon ( \rho_\phi +
p_\phi) \,.
\ee
If dissipation occurs into light degrees of freedom which quickly
thermalize and become radiation,  then we have $\rho_\alpha= \rho_R$ and: 
\be
\dot \rho_R + 4 H \rho_R  = \Upsilon \dot \phi^2\,. \label{eomrad}
\ee
Radiation  is not necessarily redshifted away during inflation,
because it is continuously fed by the inflaton 
through the dissipation \cite{Berera:1996fm}. 
Inflation happens when $\rho_R \ll \rho_\phi$, but even if small when
compared to the inflaton energy density it can be larger than the
expansion rate with $\rho_R^{1/4} > H$. Assuming thermalization, this
translates roughly  into $T > H$. This is the condition for warm
inflation, i.e., for the dissipation potentially affecting both the
background inflaton dynamics, and the primordial spectrum of the field
fluctuations. In the presence of a thermal bath, when $T > H$ the quantum
fluctuations of the fields are dominated by the thermal fluctuations,
with the amplitude of the super-Hubble fluctuations being dependent
on $T$ \cite{Berera:1995wh,Berera:1999ws}, 
therefore larger than the vacuum fluctuation with an amplitude
dependent only on $H$. Otherwise, when $T < H$ (or
similarly when $\rho_R^{1/4} < H$), one just recovers the standard cold
inflation scenario, where dissipation can  be neglected.

During warm inflation the motion of the inflaton field has to be
overdamped in order to have the accelerated expansion, but now this
can be achieved due to 
the extra friction term $\Upsilon$ instead of that of the Hubble
rate. And once $\phi$, $H$, and also $\Upsilon$, are in this slow-regime, the
same will happen with the radiation energy density, the source term
now compensating for the Hubble dilution. The equations of motion
reduce then to:
\bea
3 H ( 1 + Q ) \dot \phi &\simeq&  -V_\phi    \,,\label{eominfsl} \\
4 \rho_R  &\simeq& 3 Q\dot \phi^2\,. \label{eomradsl}
\eea
where we have introduced the dissipative ratio $Q=\Upsilon/(3 H)$. 
Notice that
 $Q$ is not necessarily constant. The
coefficient  $\Upsilon$ will depend on  $\phi$ and $T$, and 
therefore depending on the model the ratio $Q$ may increase or
decrease during inflation. In the former case, the radiation density
may also slowly increase.   
Introducing another slow-roll parameter to take into account the variation of
$\Upsilon$:
\be
\beta_\Upsilon = m_P^2 \left ( \frac{\Upsilon_\phi V_\phi }
     {\Upsilon V}\right) \,, \label{beta}
\ee
the slow-roll conditions are now given by \cite{tb,hmb,Moss:2008yb}: 
\be
\epsilon_\phi < (1 +Q) \,,\;\;\;\;\;\;       
\eta_\phi < (1 +Q) \,,\;\;\;\;\;\;       
\beta_\Upsilon < (1 +Q) \,, \label{wslowroll}
\ee        
where the condition on $\beta_\Upsilon$ ensures that the variation of
$\Upsilon$ w.r.t. $\phi$ is slow enough. Similarly, taking also into
account the dependence on $T$, we have the condition:
\be
\left|\frac{d \ln \Upsilon}{d \ln T}\right| < 4 \,,
\ee
which reflects the fact that radiation  has to be produced at a rate
larger than the redshift due to the expansion of the universe.   
In addition, given that inflation may take place in the presence of a
thermal bath, we have to worry about the thermal corrections to the
inflaton potential  not being too large, mainly that they do not
induce a too large thermal mass for the inflaton. This translates into
the condition: 
\be
\delta = \frac{T V_{T\phi}}{V_\phi} < 1  \label{delta}\,.
\ee
Once the above condition is fulfilled, the slow-roll conditions in Eq.
(\ref{wslowroll}) ensure  that
Eqs. (\ref{eominfsl}) and (\ref{eomradsl}) hold and that $\rho_R$ is
subdominant during inflation. On the other hand, when  they are not
longer satisfied, either the
motion is no longer overdamped and slow-roll ends, or the radiation
becomes comparable to the inflaton energy density. Either way,
inflation will  end shortly  afterwards. 

Given a particular model and the dissipative coefficient $\Upsilon$,
we can test the viability of warm inflation scenario by
integrating the slow-roll evolution equations, checking under which
conditions we can get enough inflation, say among 40-60 e-folds of
inflation, and checking the predictions related to the primordial spectrum.
For warm inflation we need $T >  H$, i.e., $\rho_R^{1/4} > H$,
although with $\rho_R < \rho_\phi$. But depending on the ratio $Q$ we
can have different kind of scenarios: 

(a) when $Q < 1$, dissipation is
not strong enough to affect the inflaton evolution, and we recover its
standard slow-roll EOM; still the thermal fluctuations of the
radiation energy density will modify the field fluctuations, and
affect the primordial spectrum of perturbations. This is called
{\it weak dissipative warm inflation} (WDWI). 

(b) When $Q >1$, dissipation dominates
both the background dynamics and the fluctuations, and we are in
{\it strong dissipative warm inflation} (SDWI). 
Field potentials that are not flat
enough to allow the standard slow-roll inflaton evolution can lead to
a period of inflation due to the extra friction induced by
$\Upsilon$. 

Given that the ratio $Q$ will also evolve during
inflation, we may have also models where we start say in WDWI but
end in SDWI, or the other way round.

The main ingredient now in order to study the different possibilities
is to set the functional dependence of the
dissipative coefficient on $\phi$ and $T$. 
By now there are many phenomenological models of 
warm inflation \cite{Berera:1996fm,tb,hmb,wipheno}.
More interesting are the first principles models
of warm inflation in which the dissipative coefficient
$\Upsilon$ and effective potential are computed from quantum field theory.
In particular, we will be
using the near-equilibrium approximation
proposed in \cite{BGR} with explicit expressions for $\Upsilon$
developed in \cite{mx,Berera:2008ar}. The specific field theory models 
considered for the inflaton interactions leading to dissipation 
all follow from the two-stage mechanism in \cite{BR1,br,br05}.
In this mechanism,  
the inflaton field  $\phi$ 
is coupled to heavy Bose $\chi$ and Fermi $\psi_{\chi}$ 
``catalyst'' fields, which in turn are
coupled to light fields $y_i$, $\psi_{y_i}$. 
As the background inflaton field moves down the
potential, it excites the heavy catalyst fields
which in turn decay into light degrees of
freedom \cite{BR1,br,br05,Hall:2004zr,Berera:2007qm,mossgra}. 
Consistency of the approximations then demands the
microphysical dynamics determining $\Upsilon$ to be faster than that
of the macroscopic motion of the background inflaton and the
expansion \cite{BGR}. 
Therefore, if $\Gamma_\chi$ is the decay width of the heavy
bosonic field, we need $\Gamma_\chi > |{\dot \phi}/\phi|,\, H$. In
addition, the condition $T \gg H$ allows  to neglect the expansion of 
the universe  when computing $\Upsilon$. An important feature of such
coupling configurations is that even for large perturbative couplings,
in supersymmetric (susy) models, the quantum corrections to the effective
potential from these terms can be controlled enough
to maintain an adequately flat inflaton 
potential \cite{br05,Hall:2004zr}, yet susy provides no cancellation
of the time nonlocal terms, which are the dissipative terms, and so
such terms can be quite significant. 

We will therefore consider supersymmetric models with the interactions
given in the superpotential:    
\be
W= W(\Phi)+g \Phi X^2 + h X Y^2  \label{superpot} \,,
\ee
where $\Phi$, $X$, and $Y$ denote superfields. 
We denote by $\chi$ and $y$ the bosonic components for $X$ and
$Y$ respectively, but use also $\Phi$ for the inflaton scalar
component when needed.  The background value of the inflaton field is
given then by\footnote{Without lost of generality, we will set the vev 
(vacuum expectation value) of the imaginary part of the inflaton  
field to zero.}  $\phi=\sqrt{2} \langle | \Phi | \rangle$, 
and we use $\delta \phi$ for its fluctuations. 
The full superpotential will also include the remaining interactions of $X$
and $Y$ with other fields. The latter are not relevant in order to
compute the dominant contribution to the dissipative coefficient, and
we will not specify them. For the inflaton
superpotential $W(\Phi)$ we take:
\be
W(\Phi)=\frac{\lambda}{p+1} \frac{\Phi^{p+1}}{m_P^{p-2}} \,,
\label{WPhi}
\ee
which covers the different examples given in this article: when $p
> 1$ we recover the chaotic models of inflation, and with $p=0$,
$\lambda <0$ we have the standard supersymmetric hybrid inflation. 
The potential is then given by: 
\bea
V&=& \lambda^2 m_P^{4-2p} |\frac{\phi}{\sqrt{2}}+ \delta \phi|^{2p}+ 2 \lambda g 
m_P^{2-p} {\rm Re}[ (\frac{\phi}{\sqrt{2}}+ \delta \phi)^p
  \chi^{*2}] + 4 g^2 | \frac{\phi}{\sqrt{2}} + \delta \phi|^2 |\chi|^2 + 4 h g {\rm Re}[(\frac{\phi}{\sqrt{2}}  +\delta \phi) \chi y^{*2}] \nonumber
\\ 
&& + g^2 |\chi|^4 + h^2 |y|^4 + 4 h^2 |\chi|^2 |y|^2 \,.
\eea
During inflation, the field $y$ and its fermionic partner $\psi_y$
remain massless, while the mediating field $\chi$ gets its mass from
the interaction with the inflaton field $\phi$, with 
\bea
m_{\chi_R}^2 &=& 2 g^2 \phi^2 + 2 \lambda g m_P^{2}
\left(\frac{\phi}{\sqrt{2}m_P}\right)^p \,, \label{mchiR}\\
m_{\chi_I}^2 &=& 2 g^2 \phi^2 - 2 \lambda g m_P^{2}
\left(\frac{\phi}{\sqrt{2}m_P}\right)^p \,, \label{mchiI}\\
\tilde m_{\chi}^2 &=& 2 g^2 \phi^2 \,, \label{mchiF}
\eea
$\tilde m_\chi$ being the mass of the fermionic component. 
The bosonic interactions among $\chi$, $y$ and $\delta \phi$, up to
second order in the inflaton fluctuations,  are 
given by:
\bea
V_I&=& 4 g^2 \frac{\phi}{\sqrt{2}} ( \delta \phi + \delta \phi^*)|\chi|^2 
+ 4 g^2 |\delta \phi|^2 |\chi|^2 
+ 4 h g (\frac{\phi}{\sqrt{2}}) {\rm Re}[ \chi y^{*2}] 
+ 4 h g {\rm Re}[ \delta \phi \chi y^{*2}] \nonumber \\
& & + 2 p \lambda g m_P \left(\frac{\phi}{\sqrt{2}m_P}\right)^{p-1} {\rm Re}[\delta \phi  \chi^{*2}] + 
 p (p-1) \lambda g  \left(\frac{\phi}{\sqrt{2}m _P}\right)^{p-2} {\rm Re}[\delta \phi^2  \chi^{*2}] +\cdots\label{VI}
\eea
Generically the inflationary constraints on the amount of inflation
and the amplitude of the primordial spectrum will translate in having
a small enough coupling $\lambda \ll g\,,h$, and $\lambda \ll \phi/m_P$,   
so that the terms in the second line of Eq. (\ref{VI}) can be neglected. The
dissipative coefficient arising from this pattern of interactions
among the scalar components has
been computed in Ref. \cite{mx}, and in particular in the low-$T$  
regime it is well approximated by:
\be
 \Upsilon \simeq 0.64 \times g^2 h^4 \sum_{i=R,I}\left(\frac{ g
   \phi}{m_{\chi_i}}\right)^4  \frac{T^3}{m_{\chi_i}^2} \,.     
\label{upsilon0}
\ee
The fermionic decay of the $X$ field into the $Y$ fermions, and/or
the interaction of the field fluctuations with the $X$ fermions, which
can be found in Ref. \cite{mx},  also contribute to the dissipative
coefficient, but this is subdominant in the low-$T$ regime and it can
be neglected. 
And also in general we will have $m_{\chi_R} \simeq m_{\chi_I} \simeq
m_\chi$, either due to the smalleness of the coupling $\lambda$ in
chaotic models, or whenever we are not close to the critical value in
hybrid models. The low-$T$ approximation only applies when $T/m_\chi
\lesssim 0.1$, and hereon we will restrict the analyses to this
regime.  The main reason is to be able to study the system analytically
and get the main features of the evolution.  
For models where this low-$T$ condition is
violated during inflation, we need to use the  exact
integral expressions for the dissipative coefficient given in
Ref. \cite{mx} and numerically integrate the equations. Taking into
account dissipation in the intermediate/high $T$ regime will add some
extra e-folds of inflation either at the beginning or the end of the
low-$T$ regime, depending on the evolution of $m_\chi/T$. Disregarding
this contribution what we obtain is a conservative bound on the
parameter models for warm inflation.   
On the other hand, the low-$T$ requirement $T/m_\chi \lesssim 0.1$
together with the warm inflation condition $T/H > 1$  also guarantee
that the adiabaticity condition $\Gamma_\chi > |\dot \phi/\phi|,\, H$
holds in this regime.  
With $\Gamma_\chi \simeq h^2 m_\chi/(8 \pi)$, and given that $|\dot
\phi/\phi| \simeq H|\eta_\phi/(1+Q)|$  we have then:
\be
\frac{\Gamma_\chi}{|\dot \phi/\phi|} > \frac{\Gamma_\chi}{H} >
\left(\frac{\Gamma_\chi}{m_\chi}
\right)\left(\frac{m_\chi}{T}\right)\left( \frac{T}{H} \right) >1 \label{thermalcond}\,, 
\ee
for values of the coupling $h^2 \approx O(1)$.  
This is the minimum requirement in order to have particle production of the
light degrees of freedom. Once produced, their thermalization depends
on details of their interactions with other particles, not included in
the superpotential Eq. (\ref{superpot}). We only assume that their
couplings are large enough for the thermalization to happen. 
Finally, the low-$T$ condition $m_\chi > T$ helps in keeping the thermal
corrections to the inflaton potential negligible
\cite{Hall:2004zr}.

The dissipative coefficient given in Eq. (\ref{upsilon0}) does not
depend on the coupling $g$ in this regime, but only on  
$h$ which can be rather large, $h \simeq O(1)$. The numerical
coefficient in Eq. (\ref{upsilon0}) was computed taking $X$, $Y$ to be
singlet complex fields. But in principle, they may belong to larger
representations  of a Grand Unification Theory (GUT) group, as it is
typically assumed in susy hybrid models. This will give rise to an
extra factor of ${\cal N}=N_\chi N_{decay}^2$ in front of the
dissipative coefficient, where $N_\chi$ is the multiplicity of the $X$
superfield, and $N_{decay}$ counts the no. of decay channels available
in $X$'s  decays.  Taking $m_\chi^2 = 2 g^2 \phi^2$, we have then: 
\be
 \Upsilon \simeq C_\phi \frac{T^3}{\phi^2} \,,
\label{upsilon}
\ee
where $C_\phi =0.16\times h^4 {\cal N}$.

Having set the functional form of dissipative coefficient,
Eq. (\ref{upsilon}), we can  obtain now from the slow-roll EOMs,
Eqs. (\ref{eominfsl}) and  (\ref{eomradsl}), the evolution of the
inflaton field and the ratio $Q$. From Eq. (\ref{eomradsl}) we
have the relation between $Q$ and $\phi$: 
\be
Q^{1/3}  (1 + Q )^2 = 2 \epsilon_\phi \left(\frac{C_\phi}{3}\right)^{1/3} 
\left(\frac{C_\phi}{4 C_R}\right)
\left(\frac{H}{\phi}\right)^{8/3} \left(\frac{m_P}{H}\right)^{2}
\, \label{qphi}
\ee
where $C_R= \rho_R/T^4= \pi^2
g_*/30$, with $g_*$ being the effective number of light degrees of
freedom\footnote{The different predictions and the parameter space
  available for warm inflation depend only mildy on $g_*$, and
  therefore we will take in the following the Minimal Supersymmetric
  Standard Model value $g_* \simeq 228.75$.}. 
And from Eqs. (\ref{qphi}) and (\ref{eominfsl}) we obtain the EOM:
\bea
\frac{d \phi/m_P}{d N_e} &=& -
\left(\frac{\phi}{m_P}\right)\frac{\sigma_\phi}{(1 +Q)}
\,, \label{dphine}\\   
\frac{d Q}{d N_e} &=&  \frac{Q}{1 + 7Q}\left( 10 \epsilon_\phi - 6
\eta_\phi + 8 \sigma_\phi\right)  \label{dqne}\,, 
\eea
where $N_e$ is the no. of e-folds during inflation, and we have
introduced for convenience the parameter:
\be
\sigma_\phi = m_P^2 \left( \frac{V_\phi/\phi}{V} \right) \,,
\ee
to give the variation of the inflaton field w.r.t the number of
e-folds.  This parameter is  related to the slow-roll parameter
$\beta_\Upsilon$ introduced in Eq. (\ref{beta}). In general for a
dissipative parameter depending on the inflaton field like $\Upsilon
\propto \phi^\alpha$  we have the slow-roll parameter $\beta_\Upsilon
= \alpha \sigma_\phi$; and in particular for the low-$T$ relation
Eq. (\ref{upsilon}) then $\beta_\Upsilon=-2 \sigma_\phi$. 

 Eqs. (\ref{dphine}), (\ref{dqne}), and (\ref{qphi}) are valid in
 both, WDWI and SDWI, and can give the transition from one to
 another regime during inflation. Similarly, we can get the evolution
 equation for the  ratio $T/H$, given by:
\be  
\frac{d \ln T/H}{d N_e} =  \frac{2}{1 + 7Q}\left( \frac{2 + 4 Q}{1 + Q} \epsilon_\phi - 
\eta_\phi + \frac{1-Q}{1+Q} \sigma_\phi\right)  \label{dTHne}\,, 
\ee
%

When $T>H$, the dominant contribution to the field spectrum are
the thermal fluctuations due to the radiation, and the amplitude of
the primordial spectrum is given by \cite{Berera:1995wh,Berera:1999ws,hmb}: 
\be
P_{\cal R}^{1/2} \simeq \left| \frac{H}{ \dot \phi}\right|
P_{\phi}^{1/2} 
\simeq \left( \frac{H}{2 \pi}\right) \left( \frac{3
  H^2}{V_\phi}\right)( 1 +Q)^{5/4}
\left(\frac{T}{H}\right)^{1/2} \label{spectrum}\,, 
\ee
where we have chosen to normalize the amplitude such that we recover
the CI value when $Q\rightarrow 0$ and $T/H \rightarrow 1$. 
Using now Eqs. (\ref{dphine}) and (\ref{dqne}) one can derive the
prediction for the spectral index:
\be
n_S-1 \simeq \frac{d \ln P_{\cal R}}{d \ln k} \simeq  \frac{d P_{\cal R}}{d N_e}
\simeq \frac{1}{1+Q} \left( -(2 - 5 A_Q)\epsilon_\phi -3 A_Q \eta_\phi 
 +(2 + 4  A_Q) \sigma_\phi \right) 
\,, \label{ns}
\ee
where 
\be
A_Q= \frac{Q}{1 +7 Q} \,. 
\ee
The above expression for the spectral index agrees with that given in
Ref. \cite{Moss:2008yb}, Eq. (36), by replacing into their expression
$b=0$ (no thermal corrections to the inflaton potential), and $c=3$,
$\beta_\Upsilon = -2 \sigma_\phi$, the corresponding values for the
low-$T$ dissipative coefficient Eq. (\ref{upsilon}). 

For the running of the spectral index we obtain:
\bea
n_S^\prime &= &(1 - n_S) \frac{Q^\prime}{1+Q} + \frac{1}{1+Q} 
\left( -(2 - 5 A_Q)\epsilon_\phi^\prime -3 A_Q \eta_\phi^\prime 
 +(2 + 4  A_Q) \sigma_\phi^\prime \right) \nonumber \\
&& +\frac{Q^\prime}{(1+Q) (1+7 Q)^2} \left( 5 \epsilon_\phi -3
 \eta_\phi + 4 \sigma_\phi\right) \,,
\eea
where ``prime'' denotes the derivative with respect to the no. of
e-folds, and 
\bea
\epsilon_\phi^\prime &=& \frac{2 \epsilon_\phi}{1+Q} (2 \epsilon_\phi -
\eta_\phi) \,, \\
\eta_\phi^\prime &=& \frac{\epsilon_\phi}{1+Q} (2 \eta_\phi -
\xi_\phi) \,, \\
\sigma_\phi^\prime &=& \frac{\sigma_\phi}{1+Q} (\sigma_\phi+ 2 \epsilon_\phi -
\eta_\phi) \,, 
\eea
where $\xi_\phi= 2 m_P^2(V_{\phi \phi \phi}/V_\phi)$. 

Although tensor perturbations are not affected by dissipation,
the ratio $r$ of the tensor-to-scalar perturbations will change due to 
 the modified amplitude of the primordial scalar spectrum, 
and we now have:
\be
r \simeq  \left( \frac{H}{T} \right)
\frac{ 16 \epsilon_\phi}{(1+Q)^{5/2}} \, ,
\label{rwi}
\ee
which is suppressed w.r.t. the CI prediction by a factor $(T/H) (1+
Q)^{5/2} > 1$. 

On the contrary, non-gaussian effects during warm
inflation \cite{nongauss1,nongauss2,iannongauss} can be rather large
when compared to the prediction in 
single field cold inflation models, which generally yield a very low value of
$f_{NL} \propto n_S-1 \lesssim 1$ \cite{nongaussci1,nongaussci2}; $f_{NL}$ is the
non-linearity parameter, defined by \cite{WMAP5}:
\be
\Phi= \Phi_L + f_{NL} \Phi_L^2 \,,
\ee
where $\Phi$ is the curvature perturbation during the matter era, and
$\Phi_L$ is its linear part. In the strong dissipative regime it was shown
in \cite{iannongauss} that entropy fluctuations during warm inflation
due to the thermal fluctuations play an important role in generating
large non-gaussian effects, with the prediction:
\be
-15\ln ( 1 + \frac{Q}{14}) - \frac{5}{2} \lesssim f_{NL} \lesssim
\frac{33}{2}\ln ( 1 + \frac{Q}{14}) - \frac{5}{2} \,,
\ee
which for a dissipative ratio $Q$ in the range from 10 to $10^6$ gives that
$|f_{NL}|$ ranges from 10 to 180. On the other hand, the third year
WMAP CMB data \cite{yadav} gives $26.9 < f_{NL} < 146.7$ at 95\% confidence
level, although the five year WMAP \cite{WMAP5} data gives $-9 <f_{NL}
< 111$ (95 \% CL). There seems to be a tendency for $f_{NL} >0$ in the
latest data, although this still will need to be confirmed by future
data, and in particular by data from Planck surveyor satellite
\cite{planck}. Nevertheless, a large positive value for $f_{NL}$ would
disfavour conventional cold inflation models, whereas the strong
dissipative regime could accommodate a large non-gaussian signal. And in
particular the current upper bound  $f_{NL} < 110$ translates into an upper
bound for the dissipative ratio $Q \lesssim 1.3 \times 10^4$, and
therefore on an upper bound on the dissipative parameter $C_\phi$ for
each model.      

In table \ref{table1} we summarize the expressions for the spectral
index, running of 
the spectral index, and tensor-to-scalar ratio $r$, in the WDWI limit
($Q \ll 1)$ and SDWI one ($Q \gg 1$),  including also the CI limit
for comparison. We notice again that these expressions hold for a
dissipative coefficient that depend on $T$ and $\phi$ like $\Upsilon
\propto T^3/ \phi^{2}$.   

\begin{table}[t]
{\large 
\begin{tabular}{|c|c|c|c|}
\hline
 & $ n_S-1 $ & $ n_S^\prime $ & $ r $ \\
\hline
CI & $ 2 \eta_\phi - 6 \epsilon_\phi $ & $  2 \eta_\phi^\prime - 6
\epsilon_\phi^\prime $ & $ 16 \epsilon_\phi $ \\
WDWI & $   2 \sigma_\phi -2 \epsilon_\phi   $ & $ 
2 \sigma_\phi^\prime -2 \epsilon_\phi^\prime  $  & $16 \displaystyle\frac{H}{T} \epsilon_\phi $ \\
SDWI & $ \displaystyle\frac{1}{7Q}(- 3 \eta_\phi + 18 \sigma_\phi - 9 \epsilon_\phi ) $ 
& \begin{tabular}{c} $ \displaystyle\frac{3}{7Q}(- 
    \eta_\phi^\prime + 6 \sigma_\phi^\prime - 3 \epsilon_\phi^\prime )$ \\ $ +(1 -n_S)
    \displaystyle\frac{Q^\prime}{Q} $ \end{tabular}  
& $ 16\displaystyle\frac{H}{T} \displaystyle\frac{\epsilon_\phi}{Q^{5/2}} $ \\
\hline
\end{tabular}
}
\caption{Predictions for the spectral index $n_S$, running of the
  spectral index $n^\prime_S$ and tensor-to-scalar ratio in terms of
  the slow-roll parameters for cold
  inflation (CI), and weak dissipative warm inflation (WDWI), and
  strong dissipative warm inflation (SDWI). \label{table1}}
\end{table}

In the next sections we will particularize these predictions to the
study of some simple models of inflation: (a) monomial potentials, or
chaotic inflation; (b) hybrid like potentials; (c) hilltop quadratic
potential. The first two kind of models have been studied in
Ref. \cite{bb4}, and we extend those results here,
whereas  hilltop warm inflation was studied in
\cite{BuenoSanchez:2008nc}. We assume that the inflationary dynamics
is driven by a single field, the inflaton, and that the curvature perturbation
is originated by its fluctuations. Using $C_\phi$ as a parameter, we 
search the parameter space available for each model to realise warm
inflation, and give the predictions for the spectral index, running of
the spectral index and the tensor-to-scalar ratio. The WMAP data
\cite{WMAP5} combined with the BAO and SN data indicates that $n_S=
0.96\pm 0.013$ at the
1$\sigma$ confidence level when no tensors and no running is taken
into account. Allowing $r$ and $n^\prime_S$ to be free parameters one
has $n_S=1.089^{+0.070}_{-0.068}$. The upper limit on the
tensor-to-scalar ratio is $r < 
0.22$ at the 2$\sigma$ level with no running, and $r < 0.55$  when the
running is included ($r < 0.4$ from \cite{kinney08}). For the running
of the spectral index one has $n^\prime_S = -0.053 \pm 0.028$.   

\section{Warm inflation with monomial potentials}
\label{wimonomial}

We consider in this section a general chaotic inflaton potential
\cite{Linde:1983gd} like:
\be
V=V_0 \left( \frac{\phi}{m_P}\right)^n  \label{potmon}\,,
\ee
with $n >0$. In a supersymmetric theory this kind of potentials 
can be derived from the superpotential Eq. (\ref{WPhi}), with
$n=2 p$ and $V_0= \lambda^2 m_P^4/4$. The slow-roll parameters are given by:
\be
\eta_\phi= n (n-1) \left(\frac{m_P}{\phi}\right)^2 \,,\;\;\;\;
\epsilon_\phi= \frac{n}{2(n-1)}  \eta_\phi\,,\;\;\;\;
\sigma_\phi= \frac{\eta_\phi}{(n-1)}\,.
\ee
Therefore without enough dissipation, either for cold inflation
with $Q=0$ or in the weak dissipative regime with $Q \ll 1$, these
potentials leads to inflation only for values of the inflaton field 
larger than the Planck mass $m_P$.  
Inflation ends in this case when
the slow-roll conditions are no longer fulfilled, i.e., when
$\eta_\phi \simeq1$ and then $\phi_{end} \simeq \sqrt{n(n-1)} m_P$,
therefore  the value of the field at horizon crossing $N_e$
e-folds before the end is given by:
\be
\frac{\phi_*}{m_P}   \simeq \sqrt{ n (2 N_e + n-1)}  
\simeq \sqrt{ 2 n  N_e } \,. 
\ee
for a total number of $N_e$ e-folds of either CI or WDWI. 
For CI, the predictions for the spectral index, the running of the
spectral index,  and the
tensor-to-scalar ratio are given by:
\be
n_S \simeq 1 - \frac{2 +n}{2N_e} \,, \;\;\;\; 
n_S^\prime \simeq - \frac{2 +n}{2 N_e^2} \,, \;\;\;\; 
r\simeq \frac{4n}{N_e} \,,
\ee
where $N_e$ is in the range $40-60$. 
The amplitude of the scalar primordial spectrum sets 
the value of $V_0$:
\be
\frac{V_0}{m_P^4}\simeq \frac{12 \pi^2 P_{\cal R}}{(2 n N_e )^{n/2+1}} 
\,.
\ee 

\begin{figure}[t]
\centering 
\scalebox{0.4}{\includegraphics{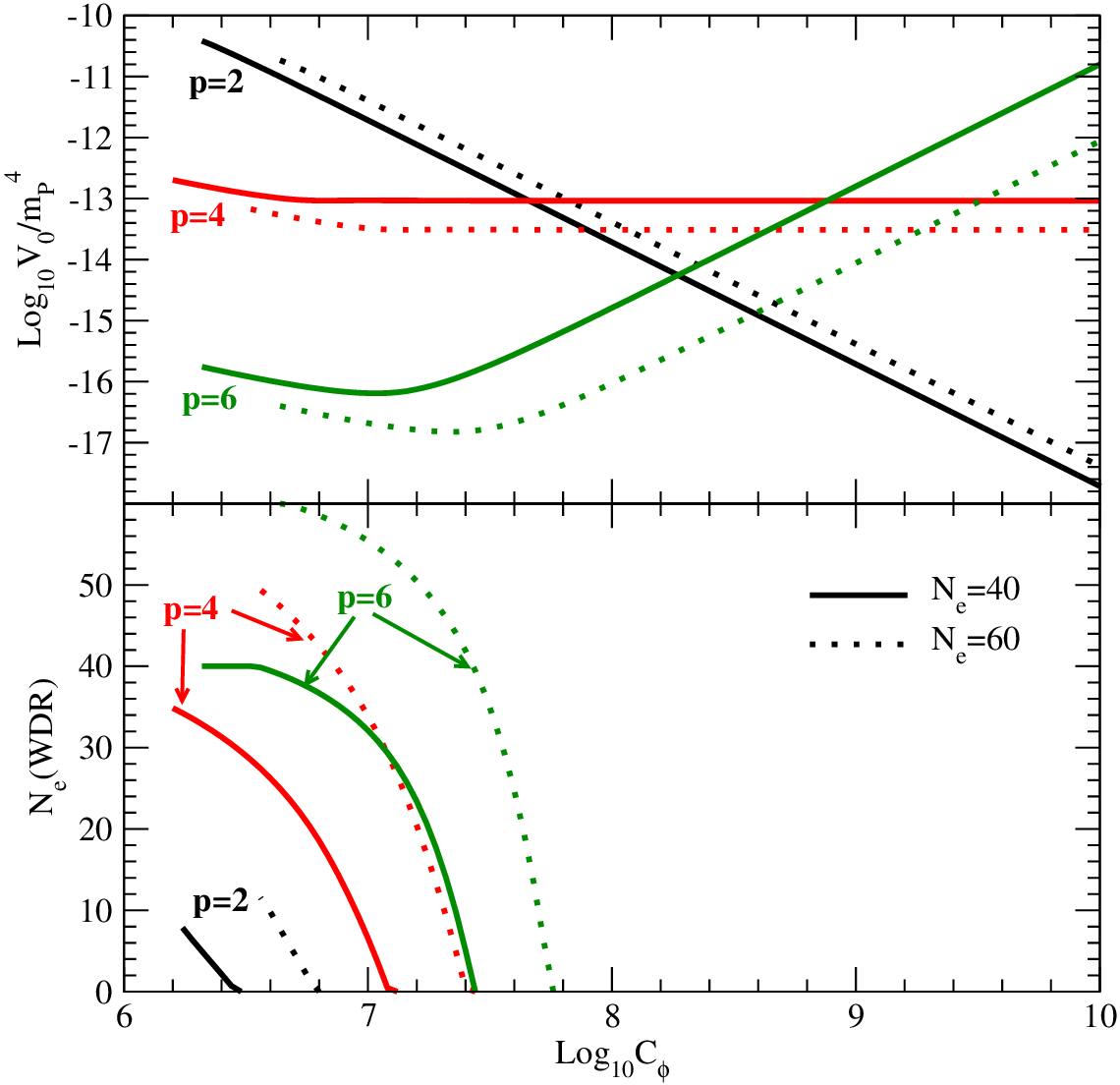}}
\caption{Top panel: Value of $V_0/m_P^4$ required to match the amplitude of the
  primordial spectrum depending on $C_\phi$, i.e, the dissipative
  parameter, for a monomial potential Eq. (\ref{potmon}) with
  $n=2,\,4,\,6$, and two choices of the total no. of 
  e-folds since horizon crossing, $N_e=40$ (solid lines), and $N_e=60$
  (dotted lines). Lower panel: no. of e-folds in WDWI, i.e., while
  $Q <1$. } 
\label{plot1}
\end{figure}

\begin{figure}[t]
\centering 
\begin{tabular}{cc}
\scalebox{0.4}{\includegraphics{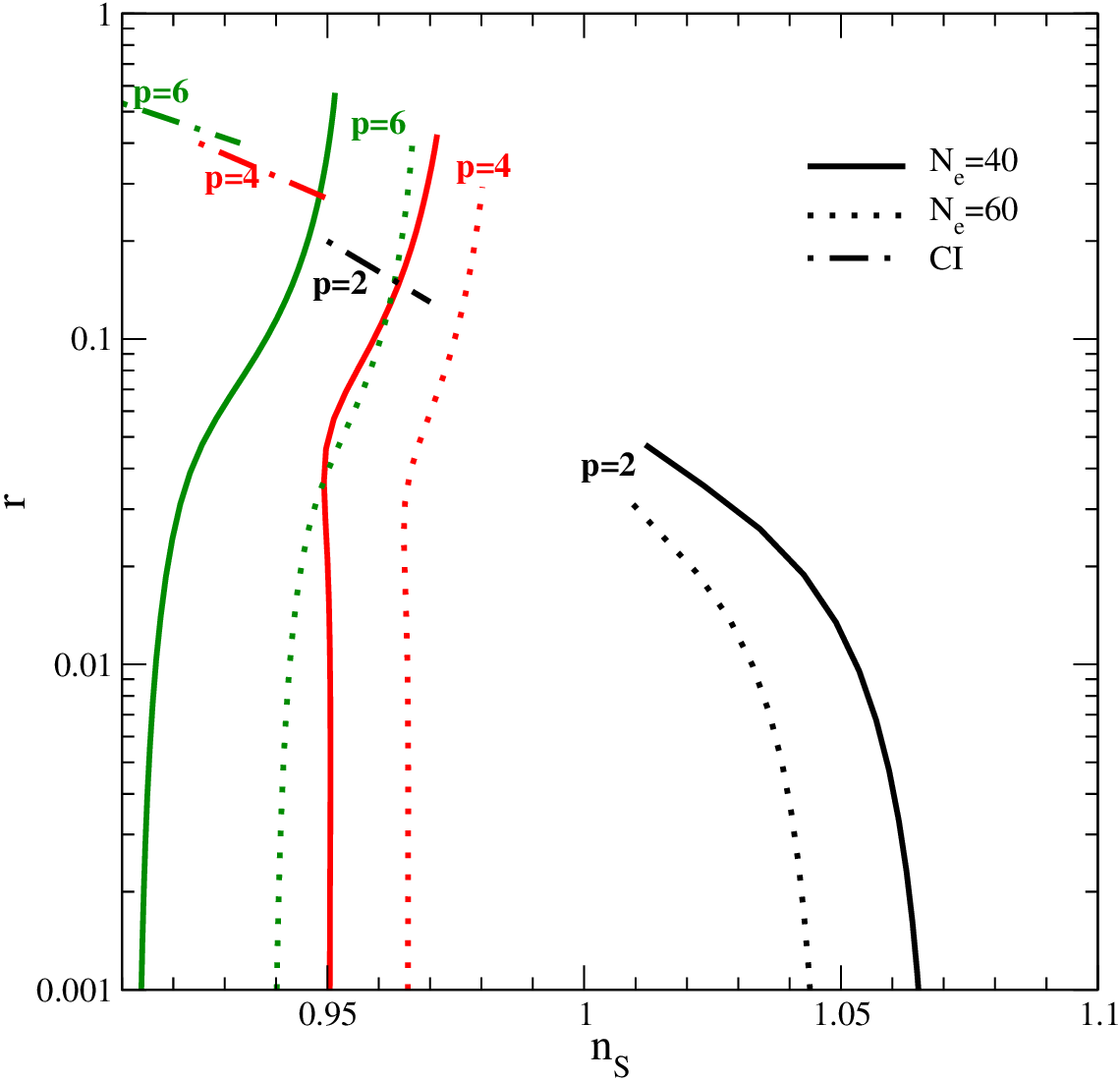}} & 
\scalebox{0.4}{\includegraphics{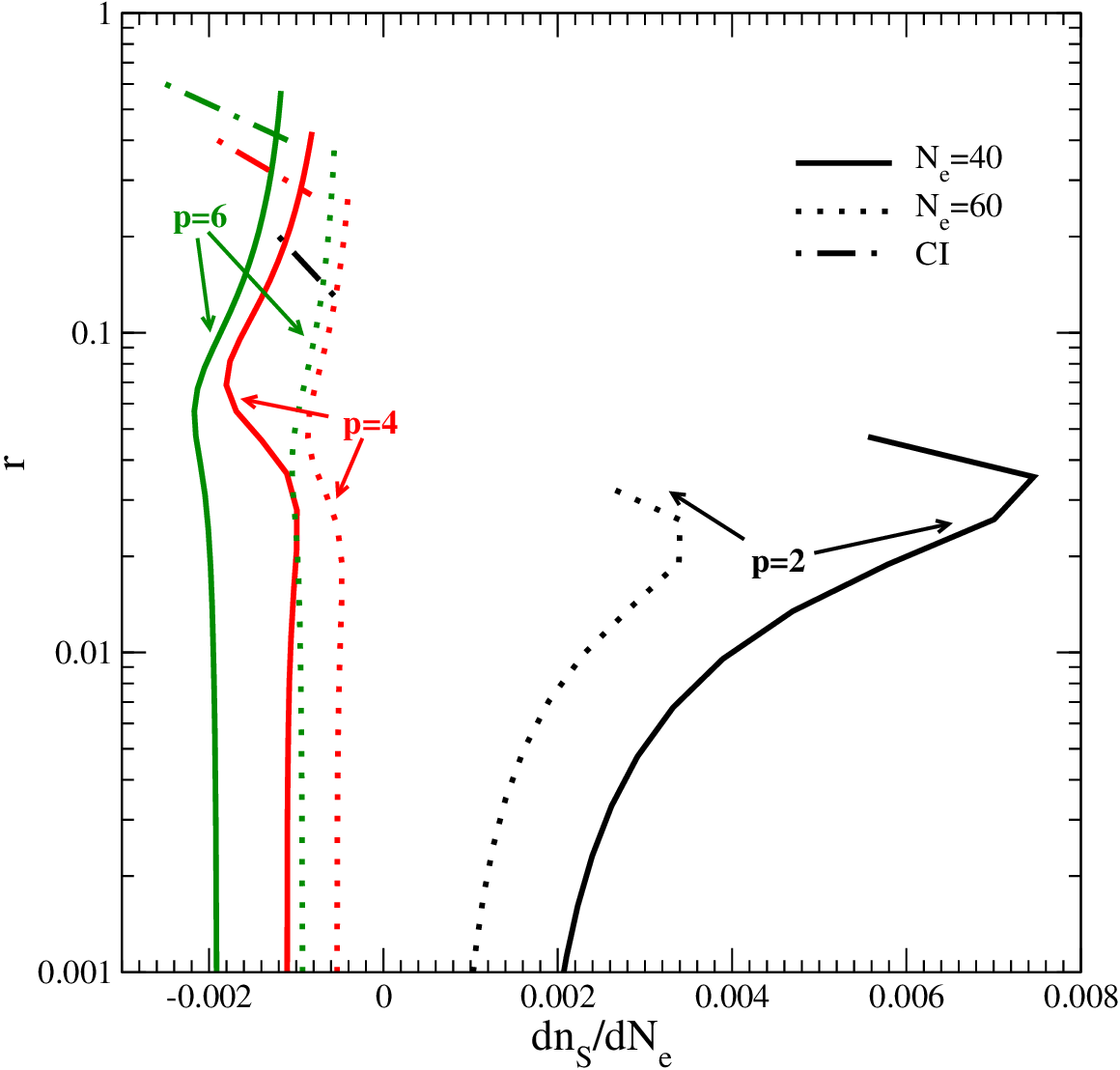}}
\end{tabular}
\caption{LHS plot: prediction of the tensor-to-scalar $r$ ratio versus the
spectral index $n_S$, for monomial potentials Eq. (\ref{potmon}) with
$n=2,\,4,\,6$, and two choices of 
the total no. of   e-folds since horizon crossing, $N_e=40$ (solid
lines), and $N_e=60$ (dotted lines). RHS plot: prediction of the
running of the spectral index $n_S^\prime$. The CI predictions (dashed
lines) are also included for comparison. }
\label{plot2}
\end{figure}

For warm inflation, the predictions will depend on the value of the
coefficient $C_\phi$, i.e., on whether we have enough dissipation
at least to keep the ratio $T/H > 1$. The inflaton field always decreases
during inflation, and the evolution of $Q$,
Eq. (\ref{dqne}), is given now by:
\be
\frac{d Q}{d N_e} =  \frac{Q}{1 + 7Q}
\frac{(14-n)}{(n-1)} \eta_\phi\,,
\ee 
and therefore the dissipative ratio increases unless $n > 14$. 
Similarly, we have:
\be
\left(\frac{T}{H} \right)^{3} = \frac{9 Q}{C_\phi}
\left(\frac{m_P^4}{V_0}\right)
\left( \frac{m_P}{\phi}\right)^{n-2} \,,
\label{thmon}
\ee
and for $n\geq 2$ the ratio $T/H$ also increases during
inflation, and thus it is enough to check for the consistency of the
scenario that this ratio is larger than one only at the beginning. 
The other condition we need to cross-check is the low-$T$
approximation $m_\chi/T = \sqrt{2} g \phi/T \gtrsim 10$. Given that
this is the only ratio that depends on the coupling $g$, instead of
varying this parameter we will
impose this condition by taking $g \sim O(1)$ and then simply $\phi/T
\gtrsim 10$, and check afterwards the constraint on the coupling $g$,
i.e., how small the coupling can be for the results to hold. 
Using Eqs. (\ref{dphine}) and
(\ref{dqne}) we have that the ratio $\phi/T$ evolves like:
\be
\frac{d \ln \phi/T}{d Ne}= \frac{1}{1+Q}
\frac{\eta_\phi}{2(n-1)}\left( n -10 - \frac{4Q }{ 1 +7Q} (14 -n)
\right) \,,
\ee
so that in WDWI with $Q \ll 1$ this ratio decreases for $n <10$,
and in SDWI it does decrease for $n<5$.    
However we have checked that as far as we fulfil the other
constraints, this ratio is always larger than 10. And for the results
shown in Figs. (\ref{plot1}) and (\ref{plot2}) we have then  that
the low-$T$ condition $m_\chi/T \gtrsim 10$ holds for $n=2$ with $g
\gtrsim 0.01$, for $n=4$ with $g \gtrsim 10^{-4}$, and for $n=6$ with $g
\gtrsim 4 \times 10^{-5}$. 

The slow-roll parameter $\eta_\phi/(1+Q)$ evolves like:
\be
\frac{d }{dN_e} \left(\frac{\eta_\phi}{1+Q}\right)= \left(\frac{\eta_\phi}{1+Q}\right)^2
\left( \frac{ 2 + Q n}{1+ 7Q} \right) \frac{1}{n-1} \,,
\ee
and therefore it increases during both WD and SD warm inflation.  
Consequently, similarly to CI, inflation ends  when $\eta_\phi =1+
Q_f$, with $Q_f$  being the value at  
the end. We remark that the slow-roll 
conditions are always violated before the energy density in radiation
dominates. Using Eqs. (\ref{qphi}) (see below Eq. (\ref{qphimon})) and
(\ref{thmon}) we have the relation:
\be
\frac{\eta_\phi}{1+Q} = \left(\frac{4 (n-1)}{n}\right)
\left(\frac{\rho_R}{V}\right)  \left(\frac{1 + Q}{Q}\right) \,,
\ee  
so that even when $Q_f \gg 1$ we still have $\rho_R \lesssim V$. 

\begin{table}[b]
{\large 
\begin{tabular}{|c|c|c|}
\hline
 & $ n_S-1 $ & $ n_S^\prime $  \\
\hline
CI & $ -n (n+2) \left(\displaystyle\frac{m_P}{\phi_*}\right)^2$ & $-2 n^2 (n+2) \left(\displaystyle\frac{m_P}{\phi_*}\right)^4$ \\
WDWI & $-n (n-2) \left(\displaystyle\frac{m_P}{\phi_*}\right)^2$ & $ -2 n^2 (n-2) \left(\displaystyle\frac{m_P}{\phi_*}\right)^4$ \\
SDWI & $\displaystyle\frac{3}{14Q_*} (14 -5 n) \left(\displaystyle\frac{m_P}{\phi_*}\right)^2 $ 
&  $\displaystyle\frac{3 n^3}{98 Q_*^2} (14 -5 n) \left(\displaystyle\frac{m_P}{\phi_*}\right)^4 $  \\
\hline
\end{tabular}
}
\caption{Predictions for the spectral index $n_S$ and  running of the
  spectral index $n^\prime_S$ for monomial potentials
  Eq. (\ref{potmon}), for cold 
  inflation (CI), weak dissipative warm inflation (WDWI), and
  strong dissipative warm inflation (SDWI). \label{table2}}
\end{table}

The procedure to get the predictions for the spectral index, ratio $r$
and the running of the spectral index is as follows: 
given the value of $Q_f$, for each value of $C_\phi$ we can integrate
Eq. (\ref{dqne})  back $N_e$ e-folds in order to get the value of the
dissipative ratio at horizon crossing $Q_*$; the value of $V_0$ is
derived by matching the amplitude of the 
primordial spectrum with the observational value  $P_{\cal R}^{1/2}
\simeq 5 \times 10^{-5}$ \cite{WMAP5}. Mainly we are
tracing the time evolution during inflation in terms of the 
the dissipative ratio $Q$, getting afterwards the value of $\phi$ from
Eq. (\ref{qphi}) , which for monomial potentials is
given by:
\be
Q^{1/3}  (1 + Q )^2 = n^2 \left(\frac{C_\phi}{3}\right)^{1/3}
\left(\frac{C_\phi}{4 C_R}\right)  
\left(\frac{V_0}{3 m_P^4}\right)^{1/3} \left(\frac{m_P}{\phi}\right)^{(14-n)/3}
\, . \label{qphimon}
\ee
The evolution Eq. for $Q$ is then:  
\be
\frac{d Q}{d N_e}= C_Q \frac{Q^{1+2s} (1 +Q)^{12s}}{1+7 Q} \,,
\label{dqnemon}
\ee
with
\bea
C_Q&=& n (14 -n) \left(\frac{3^{1/3}~2}{D_Q n^2} \right)^{6s}
\left(\frac{V_0}{m_P^4}\right)^{2s} \,,\\
s &=&  \frac{1}{14-n} \,,
\eea
And the amplitude of the spectrum reads in this case:
\bea
P_{\cal R}^{1/2} &=& C_{P} Q_*^{(2-n)s/n} (1+ Q_*)^{(38-13 n)s/4} \,,\\
C_{P}&=& \frac{1}{2 \pi} \left( 3 p^6 C_\phi\right)^{(n-2)s/2} 
\left( \frac{C_\phi}{12 C_R}\right)^{(4+n)s} 
\left( \frac{V_0}{m_P^4}\right)^{6s} \,.
\eea

Depending on the value of $C_\phi$, we can have different scenarios with:
(i) $N_e$ e-folds in 
WDWI, i.e., $Q_*,\,Q_f <1$; (ii) $Q_*> 1$ already at
$N_e$, i.e., SDWI; (iii) a transition after some e-folds from
WDWI to SDWI, starting with $Q_* < 1$ but with $Q_f >1$. 
In terms of the values for $\phi_*/m_P$ and $Q_*$, the expressions
for the spectral index and the running of the
spectral index, for a general power $n$ , are given in table
\ref{table2}, where we also include 
those from CI for comparison. The tilt in the spectrum and
the running are also negative for warm inflation with $n >2$, although 
the tilt is smaller. However for a quadratic potential with $n=2$, when
horizon crossing takes place well in WDWI with $Q_* \ll 1$, at
first order in the slow-roll parameters one has a scale-invariant
spectrum with no tilt and no running. Although this would be modified
when including higher order corrections in the slow-roll
parameters. In SDWI on the contrary, for a quadratic potential we
have a blue-tilted spectrum with positive running.

Nevertheless, unless $n>6$, we do not get 40-60 e-folds only in
WDWI. Depending on the value of the parameter $C_\phi$, for the smaller values
inflation begins in WDWI, but after some e-folds, as $Q$ increases,
it enters into SDWI where inflation ends. This can be seen in the
lower panel of Fig. (\ref{plot1}), where is plotted the no. of e-folds
of expansion in WDWI since horizon crossing, assuming that the
total no. of e-folds is either 40 (solid line) or 60 (dotted line). On
the top panel we have included the value of $V_0$ in units of $m_P^4$
required to get the amplitude of the primordial spectrum, i.e., the
value of the coupling $\lambda$. For example for $n=2$ this coupling
is below $10^{-5}$, decreasing when increasing $C_\phi$; while for
$n=4$ is rather constant with $C_\phi$ and of the order of $6\times 
10^{-7}$. Therefore, as mentioned in the previous section, unless the
coupling $g$  is also that small, the bosonic interactions induced by
$\lambda$ gives a negligible contribution to the dissipative
coefficient.  

The lower limit on $C_\phi$ is given by the condition $T/H \geq 1$,
with $Q_*$ and $V_0$ consistent with the amplitude of the primordial
spectrum. The upper limit would be given by demanding not to exceed
the current limit on non-gaussianity, i.e, $|f_{NL}| < 110$, which
gives $Q_* < 1.3\times 10^4$. The latter translates into $C_\phi
\lesssim 2.5\times 10^8$ ($5.2\times 10^8$) for $n=2$ and $Ne=40\,
(60)$, $C_\phi \lesssim 1.1\times 10^9$ ($2.3\times 10^9$) for $n=4$,
and  $C_\phi \lesssim 2.5\times 10^9$ ($5.2\times 10^9$) for $n=6$. 
The dependence on $C_\phi$ for the different parameters and predictions is
summarized in Table \ref{table3}.  

\begin{table}[b]
{\large 
\begin{tabular}{|c|c|c|c|c|c|c|}
\hline
 & $V_0/m_P^4$ & $\phi_*/m_P$ & $Q_*$ & $ n_S-1 $ & $ n_S^\prime $ & $r$   \\
\hline
WDWI & $\propto C_\phi^{-1} $&$\propto C_\phi^{0} $&$\propto
C_\phi^{3} $&$\propto C_\phi^{0} $&$\propto C_\phi^{0} $ & $\propto C_\phi^{-3} $ \\
SDWI & $\propto C_\phi^{4-n} $&$\propto C_\phi^{-1} $&$\propto
C_\phi^{2} $&$\propto C_\phi^{0} $&$\propto C_\phi^{0}
$&$\propto C_\phi^{-7} $ \\
\hline
\end{tabular}
}
\caption{Dependence on the parameter $C_\phi$, for weak dissipative
  warm inflation (WDWI), and strong dissipative warm inflation
  (SDWI) for monomial potentials Eq. (\ref{potmon}). \label{table3}} 
\end{table}

The predictions for the spectral index, running and tensor-to-scalar
ratio are given in Fig. (\ref{plot2}). 
We have also included for comparison the  predictions obtained in CI
when varying the no. of e-folds between 40 and 60 (dashed lines). 
Well into WDWI, or SDWI, the spectral index and the running do
not depend on the parameter $C_\phi$, while $r$ quickly decreases with
$C_\phi$: in WDWI this is due to the increasing factor $T/H$, and
in SDWI we have also the effect of $Q_*$ which makes the tensor
contribution negligible with respect to the scalar one. 
Typically for warm inflation the spectrum is less tilted
compared to the cold inflation scenario (and for $n=2$ is
blue-tilted), and the running is slightly larger. These two effects,
the decrease in both $r$ and the tilt of the spectrum, makes now 
{\it the quartic potential  compatible with observations} 
\cite{WMAP5,kinney08} even having only weak dissipation at horizon
crossing. For example with 
$C_\phi \sim 6 \times 10^{6}$ we have $n_S\simeq 0.96 ~(0.98)$ and
$r\simeq 0.13~ (0.235)$ for $N_e =40~(60)$; while in CI we have
$n_S\simeq 0.925 ~(0.95)$ and $r\simeq 0.4~ (0.27)$, values that seem
disfavoured by the latest data \cite{WMAP5,kinney08}. Similarly for
$p=6$, as can be seen in Fig. (\ref{plot2}). On the other hand, by
increasing the value of the dissipative parameter $C_\phi$, inflation
takes place in SDWI, the tensors are negligible, and the value of
the inflaton field at horizon crossing decreases {\it below the
  Planck mass}. For $p=2$ this  happens with $C_\phi \gtrsim 10^7
\, (2.7\times 10^7)$ and $Q_* \gtrsim 20\, (35)$ for $N_e=40$ (60);       
 for $p=4$ we have $C_\phi \gtrsim 9.1\times 10^7
\, (2.3\times 10^8)$ and $Q_* \gtrsim 94\, (137)$; and        
 for $p=6$ we have $C_\phi \gtrsim 3.3\times 10^8
\, (7.6\times 10^8)$ and $Q_* \gtrsim 234\, (282)$.         
 
\section{Hybrid models}
\label{wihybrid}

We consider now small field models of inflation, with an inflationary
potential given by:   
\be
V=V_0 \left( 1 + \frac{\gamma}{n} \left(\frac{\phi}{m_P}\right)^n
\right)\,, \label{Vn}
\ee
in general for $n > 0$, and a logarithmic potential for $n=0$:
\be
V=V_0 \left( 1 + \gamma \ln \frac{\phi}{m_P}\right)\,. \label{Vlog}
\ee
The above potentials can be regarded as the inflationary part of the
supersymmetric hybrid potential derived from Eq.(\ref{WPhi}) with
$p=0$ and $\lambda <0$, such that $V_0=\lambda^2 
m_P^4$. At tree-level the inflationary potential is a flat potential
given by $V_0$, but lifted either by the logarithmic radiative
corrections due to the mass splitting in the $X$ superfield, as in
Eq. (\ref{Vlog}), or by the 
soft susy breaking mass term contributions, like in Eq. (\ref{Vn})
with $n=2$. In the former case we have $\gamma=g^2 N_\chi/(8 \pi^2)$, $N_\chi$
being the multiplicity of $X$, while for the quadratic potential we
have simply $\gamma=m_\phi^2m_P^2/V_0=\eta_\phi$, $m_\phi$ being the
inflaton mass. By lifting the potential by adding non-renormalizable
interactions, or taking into account supergravity (sugra) corrections coming
from the K\"ahler potential \cite{sugra, sugra2, senoguz},  we can also have
Eq. (\ref{Vn}) with $n \geq 2$.  
In general in supersymmetric models we may have all kind of lifting terms
in the potential, starting with the log term from the radiative
corrections, the quadratic mass from the soft susy breaking and/or
sugra, and higher powers from sugra or non-renormalizable terms, and
depending on  the scale of inflation and the values of the couplings
the dominant term can change as inflation proceeds. As we are mainly
interested on 
the impact of the dissipative dynamics on the inflationary
predictions, we will restrict the analyses in this section to the case
when only one possible contribution dominates, in particular to the
cases $n=0,\,2$, but this can be easily generalised to higher powers.

The total energy density during inflation is dominated anyway by the
vacuum potential energy $V_0$, such that $H$ practically remains
constant during inflation. But once the inflaton reaches the critical
value for which the $\chi_R$ squared mass (Eq. (\ref{mchiR})) becomes
negative, the tachyonic instability in the $\chi$ field will drive the
system towards the global minimum of the potential with $\langle \chi
\rangle \neq 0$ and $\phi=0$ \cite{linde,shafi,Copeland:1994vg}. Inflation
may last at most until the field reaches the critical value, or may
end before when the slow-roll conditions are no longer fulfilled.   
The slow-roll parameters are given by:
\be
\eta_\phi= (n-1) \gamma \left(\frac{m_P}{\phi}\right)^{(2-n)} \,,\;\;\;\;
\epsilon_\phi= \frac{\gamma^2}{2}\left(\frac{m_P}{\phi}\right)^{(2-2
  n)}   \,,\;\;\;\; 
\sigma_\phi= \frac{\eta_\phi}{(n-1)}\,.
\ee
Inflation can take place for small field values with $\phi < m_P$ as
far as $\gamma$ is small enough. And once dissipation is taken into
account, this can be the case even for values of $\gamma$ larger than
unity but with $\gamma/(1+Q) \ll 1$. When the potential is lifted by a
mass correction, this means 
in particular that with dissipation a mass term larger than the Hubble
scale can be compatible with inflation. Such contributions can arise
in sugra models when taking into account the corrections from the
K\"ahler potential, the so-called ``eta'' problem in sugra inflation
\cite{Dine}.  
The problem can be overcome by considering some specific forms of the
K\"ahler potential which lead to the cancellation of the quadratic term
in the potential, like in the standard susy hybrid inflationary model
with minimal K\"ahler potential \cite{shafi,Copeland:1994vg} or by
working with no-scale K\"ahler potentials
\cite{Gaillard:1995az,Murayama:1993xu}; 
another alternative would be considering D-term inflation models
\cite{Dterminf}, for which the vacuum energy originated in the
potential D-term instead of the superpotential. Warm inflation
provides therefore a different solution to the problem, relaying on the
interactions of the inflaton field but without restricting the origin
of the vacuum energy and/or the kind of sugra embedding.       

The slow-roll equations of motion for warm inflation read:
\bea
\frac{d \ln \phi/m_P}{d N_e} &=& -\frac{\gamma}{1+ Q}
\left(\frac{\phi}{m_P}\right)^{n-2} \,, \label{phiNen}\\
\frac{d \ln Q}{d N_e} &=& \frac{\gamma}{1+7Q}
\left(\frac{\phi}{m_P}\right)^{n-2} \left( 14 - 6 n + 5 \gamma
\left(\frac{\phi}{m_P}\right)^n \right) \,, \label{QNen}\\
\frac{d \ln T/H}{d N_e} &=& \frac{2 \gamma}{1+7Q}
\left(\frac{\phi}{m_P}\right)^{n-2} \left(  \frac{2}{1+Q} -  n + \gamma
\frac{1 + 2Q}{1+Q}
\left(\frac{\phi}{m_P}\right)^n \right) \,, \label{THNen}\\
\frac{d \ln \phi/T}{d N_e} &=& -\frac{\gamma}{1+Q}
\left(\frac{\phi}{m_P}\right)^{n-2} \left(  1 + \frac{2}{1+7Q}( 2 - n
(1 +Q)) + \frac{\gamma}{2} \left(\frac{\phi}{m_P}\right)^n 
\frac{ 3+ Q}{1+7Q} \right) \,. \label{phiTNen}
\eea
Therefore, for both $n=0,\,2$ the dissipative ratio $Q$
increases during inflation, and we can have the three kinds of
scenarios: (a) 40-60 e-folds with weak dissipation only; (b)
Starting in WDWI but ending inflation in SDWI; (c) 40-60 e-folds
in the strong dissipative regime. On the other hand, $T/H$ always
increases for the logarithmic potential, so once the condition $T>H$
is fulfilled at the start of inflation the system will remain in the
warm regime; for the quadratic potential it does so in WDWI with $Q
<1$, but once in SDWI this ratio decreases. For the slow-roll
parameters, the ratio $\eta_\phi/(1+Q)$ in WDWI increases for
$n=0$, but it remains constant for $n=2$, while in SDWI it
decreases in both cases. Thus, in SDWI inflation will end either
when the field reaches the critical value or by violating one of the
conditions for low-$T$ dissipation, whatever happens first.

The relation between $Q$ and the vev of the inflaton field
Eq. (\ref{qphi}) is given by:   
\be
Q^{1/3} (1 + Q)^2 \simeq
\left(\frac{C_\phi}{3}\right)^{1/3}\left(\frac{C_\phi}{4 C_R}\right)
\left(\gamma (\frac{H}{m_P})^{n-2}\right)^{2}
\left(\frac{\phi}{H}\right)^{2n -14/3} \label{qphin}\,.
\ee
Notice that this relation, and therefore the different predictions,
depends on $\gamma$ through the combination  
$\tilde \gamma= \gamma (H/m_P)^{n-2}$, so we will use $\tilde \gamma$ as
a model parameter in the following. For a
quadratic potential this is just  the ratio of the
inflaton mass and the vacuum energy, $\tilde \gamma= \gamma= m^2_\phi m_P^2/V_0$,
whereas for the logarithmic potential $\tilde \gamma$ is the ratio
between the coupling constant $3 \gamma=3 g^2 N_\chi/(8 \pi^2)$ and the vacuum
energy $V_0$ in Planck units. Using the relation  Eq. (\ref{qphin}) we have
for the amplitude of the primordial spectrum during warm inflation: 
\be
P_{\cal R}^{1/2} \simeq \frac{1}{2 \pi} \left( \frac{C_\phi}{4 C_R} \right)^{1/2}\left( \frac{H}{\phi_*} \right)
(1+Q_*)^{1/4} \,, \label{PRn}
\ee
where we have left explicit the dependence on both $H/\phi_*$ and $Q_*$ for
simplicity. Indeed the spectrum only depends on one of these variables, say the
value of the dissipative ratio at horizon crossing $Q_*$, and
the model parameters $C_\phi$ and $\tilde \gamma$.  The recipe now in
order to get the parameter space available for warm inflation would be:

(a) For each pair of values ($\tilde \gamma$, $C_\phi$), get the value
of $Q_*$ ($\phi_*$) compatible with the observable value of primordial
spectrum from Eq. (\ref{PRn}). 

(b) Check with Eqs. (\ref{phiNen}) and (\ref{QNen}) whether one can
obtain $Ne = 40-60$ e-folds of slow-roll inflation while having
$T/H>1$ and $\phi/T > 10$, and $\rho_R < V_0$. The radiation energy
density during warm inflation is given by:
\be
\frac{\rho_R}{V_0} = \frac{C_R}{ 9 } \left( \frac{T}{H} \right)^4
\frac{V_0}{m_P^4} \,,
\ee
so that in principle it can be kept below $V_0$ by lowering the scale
of the potential if needed. 
 
(c) Check the predictions for the different observables: spectral
index, running of the spectral index, tensor-to-scalar ratio.

\begin{table}[b]
{\large 
\begin{tabular}{|c|c|}
\hline 
 $\;\;\;\;\;\;\;\;\;\;\;\;\;\;\;\;\;\;\;\;\;\;\;\;\;\;\;\; n_S-1 $ & $ \;\;\;\;\;n_S^\prime $  \\
\begin{tabular}{ccc}  & $n=0$ & $n=2$ \\ \hline  
CI & $-2 \gamma \left(\displaystyle\frac{m_P}{\phi_*}\right)^2$ & $2 \gamma$ \\ 
WDWI & $\gamma (2 - \gamma) \left(\displaystyle\frac{m_P}{\phi_*}\right)^2$ & $2 \gamma$ \\ 
SDWI & $3 (14 - 3 \gamma) \displaystyle\frac{\gamma}{14 Q_*} 
  \left(\displaystyle\frac{m_P}{\phi_*}\right)^2$ & $ \displaystyle\frac{15\gamma}{7
    Q_*}$ 
\end{tabular} 
&
\begin{tabular}{cc} $n=0$ & $n=2$ \\ \hline 
$-4\gamma^2 \left(\displaystyle\frac{m_P}{\phi_*}\right)^4$
  & $8\gamma^3 \left(\displaystyle\frac{\phi_*}{m_P}\right)^2$ \\ 
$4\gamma^2 \left(\displaystyle\frac{m_P}{\phi_*}\right)^4$
  & $2 \gamma^2( 2 \left(\displaystyle\frac{\phi_*}{m_P}\right)^2 - Q_*)$ \\ 
$\displaystyle\frac{2 \gamma^2}{7 Q_*^2} \left(\displaystyle\frac{m_P}{\phi_*}\right)^4(
  \displaystyle\frac{4}{Q_*}+ 3 \gamma - \displaystyle\frac{81}{28} \gamma^2)$  
  & $-\displaystyle\frac{30 \gamma^2}{49 Q_*^2} $
\end{tabular} 
\\
\hline
\end{tabular}
}
\caption{Predictions for the spectral index $n_S$ and  running of the
  spectral index $n^\prime_S$, at leading order in $\gamma$, $Q_*$,
  $\phi_*$, for the potentials given in Eqs. (\ref{Vn}) and
  (\ref{Vlog}), for cold  
  inflation (CI), weak dissipative warm inflation (WDWI), and
  strong dissipative warm inflation (SDWI). \label{table4}}
\end{table}

The expressions for the spectral index and the running at leading
order in the parameters are given in Table \ref{table4} for the
different regimes, including those for cold inflation for
comparison. We stress that when $\gamma <1$ in the warm inflationary
regime, both weak 
and strong, the spectral index is blue tilted even for the
logarithmic potential with a negative curvature of the potential. For
small couplings, the
spectral index would be blue-tilted in the weak dissipative regime
whatever the dominant power in the potential at the time of horizon crossing,
and it will only become red-tilted in the strong dissipative regime when
$n>6$. Nevertheless, for a large enough coupling the spectral index for a
logarithmic potential can be rendered red-tilted with $\gamma > 2$ in
the weak dissipative regime, and $\gamma > 14/3$ in the strong
dissipative regime. However, for the tensor-to-scalar ratio, 
taking into account
Eqs. (\ref{qphin}) and (\ref{PRn}) we recover for warm inflation the
same prediction as in cold inflation:
\be
r\simeq \frac{2}{\pi^2 P_{\cal R}} \left(\frac{H}{m_P} \right)^2
\simeq 8.14 \times 10^{-3} \left( \frac{V_0^{1/4}}{10^{16}\, {\rm
      GeV}}\right)^4 \,, \label{rVn}
\ee
for any power $n$ in the potential. Therefore for a scale of
inflation $V_0 < 10^{16}$ GeV we have a negligible tensor contribution
to the spectrum with $r < 0.01$.

\begin{figure}[t]
\centering 
\begin{tabular}{cc}
\scalebox{0.4}{\includegraphics{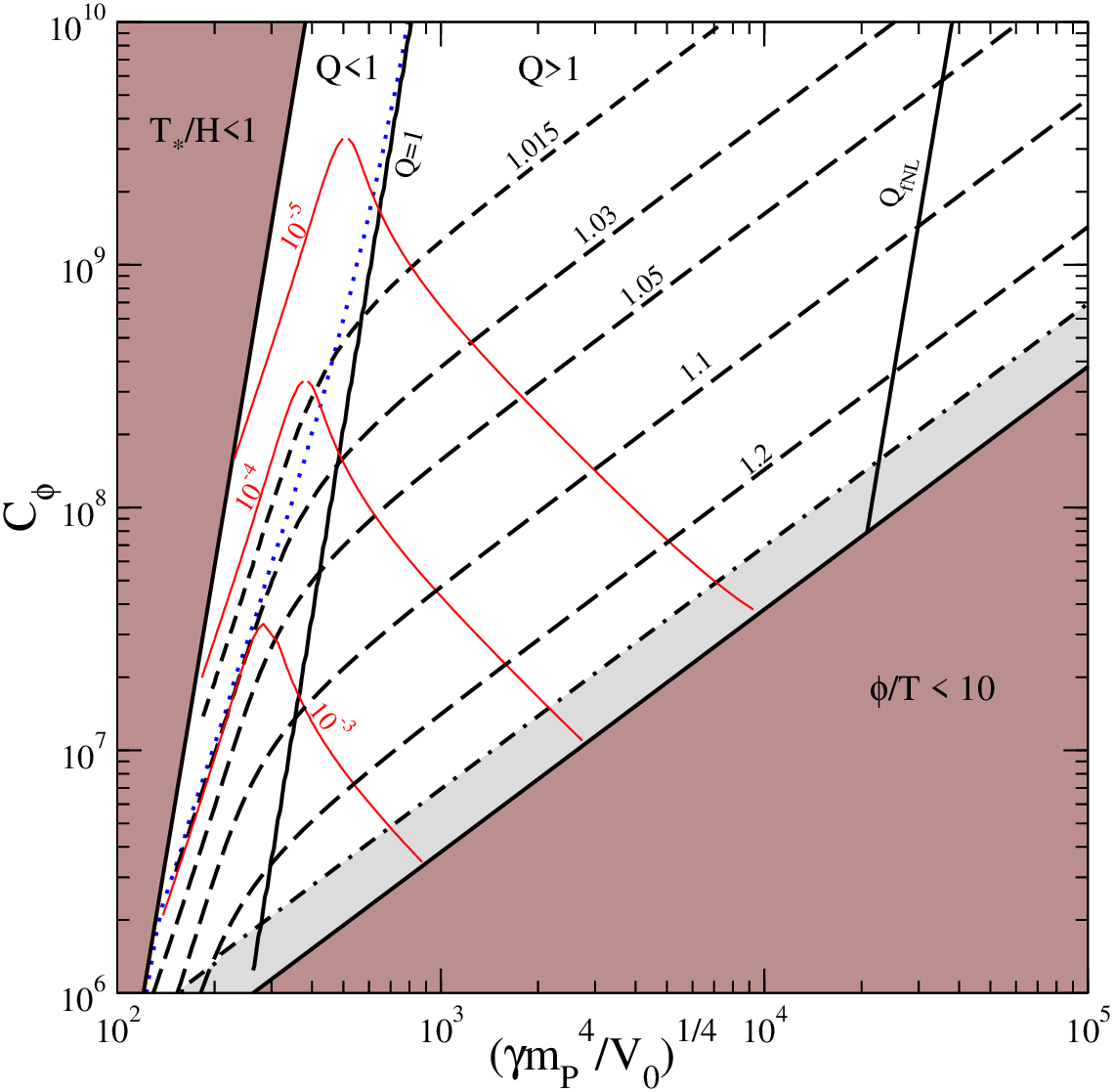}} & 
\scalebox{0.4}{\includegraphics{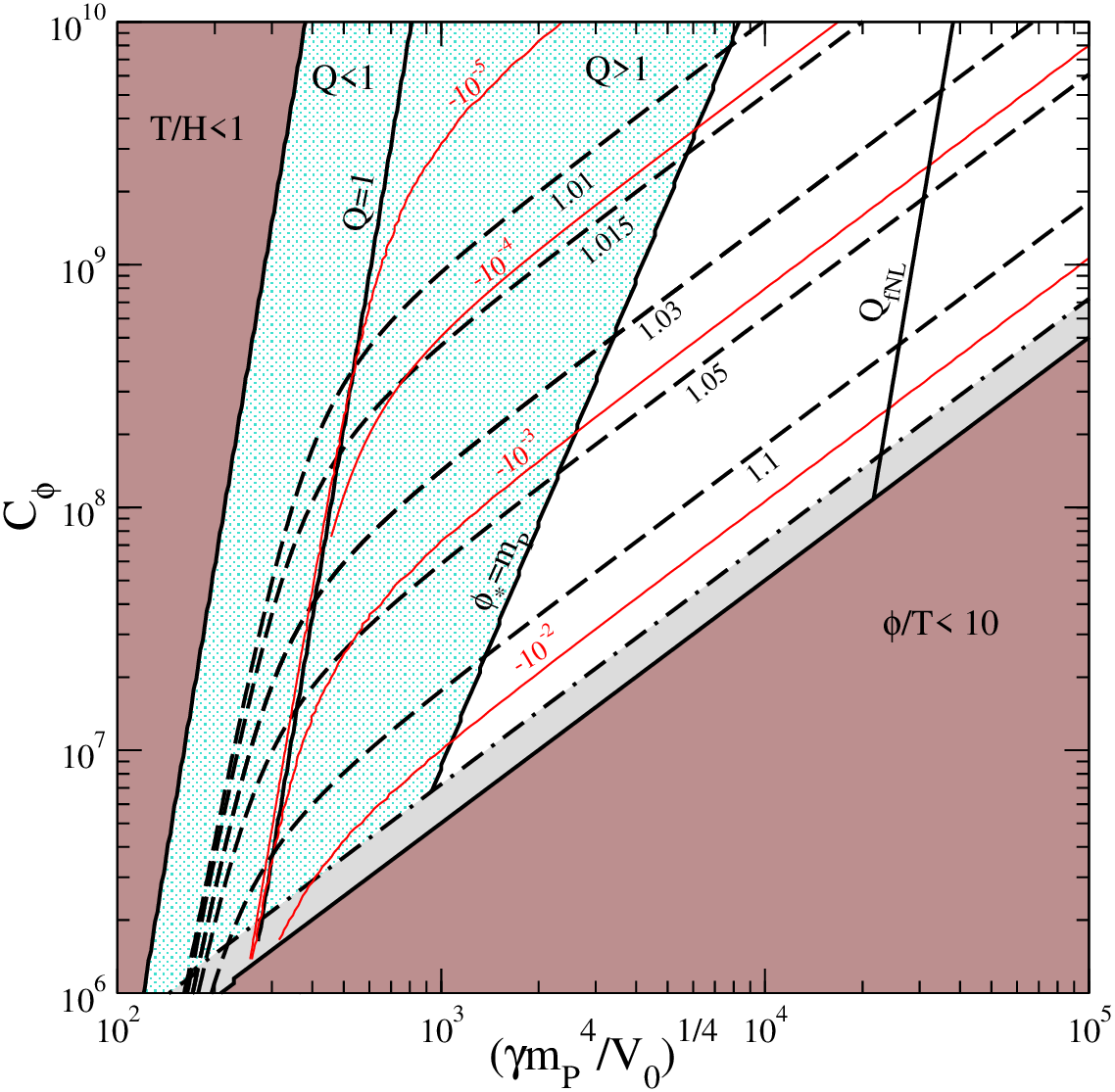}}
\end{tabular}
\caption{The $C_\phi$-$(\gamma m_P^4/V_0)^{1/4}$ plane for the
  logarithmic potential Eq. (\ref{Vlog}). Within the dark grey (brown)
  area we have $N_e < 40$ ( $N_e < 0$ within the light grey area). To
  the right of the line $Q=1$ we have strong
  dissipative warm inflation (SDWI), while to the left we have weak
  dissipative warm inflation (WDWI). 
  Between the line $Q=1$ and the dotted (blue)
  line inflation starts in WDWI but ends in SDWI. The dashed
  lines show the prediction for the spectral index, while the thin
  (red) lines that of the running of the spectral index. In the LHS
  plot we have taken $\gamma=10^{-4}$, while in the RHS we have
  $\gamma=2$. In the dotted area of the RHS plot, to the left of the line labelled $\phi_**=m_P$,  the value of the
  inflaton field at horizon crossing is larger than $m_P$.  }
\label{plotlog}
\end{figure}

 In Fig. \ref{plotlog} we have plotted the parameter space available
 to get 40 (60) e-folds of warm inflation with a log potential, in the
 plane $C_\phi$-$\tilde \gamma$, consistent with the
 observed amplitude of the primordial spectrum. On the LHS plot we have
 taken a small coupling with $\gamma=10^{-4}$, while on the RHS plot we
 have $\gamma=2$. Given than in this case during warm inflation the
 ratio $T/H$ increases but $\phi/T$ decreases, the left shaded area is
 excluded because $T/H <1$ at horizon crossing, while the right shaded
 area is excluded because we reach $\phi/T=10$ before 40 e-folds after
 horizon crossing (60 e-folds for the light grey shaded area). 
This means that roughly the low-$T$ dissipative regime ends, but still
one could get some extra e-folds in the intermediate and high-$T$
regime before inflation ends. The condition on $\phi/T$ only gives the most
conservative upper limit on $\tilde \gamma$ for a given value of
$C_\phi$. On the other hand, for a logarithmic potential a small value
of the coupling $\gamma=g^2 N_\chi/(8 \pi^2)$, means that to fulfill
the condition $m_\chi/T \gtrsim 10$  we need a larger value of the
ratio $\phi/T$. For example with $g\simeq 0.01$ one would have to
impose $\phi/T \gtrsim 10^3$ instead, and the  limit on $C_\phi$
increases approximately by a factor of 6 with respect to that shown in
Fig. (\ref{plotlog}); similarly for $g\simeq 10^{-3}$ the limit
increases approximately by a factor of 30. 

 In all the parameter space available, after 40 (60) e-folds the slow-roll
 conditions still hold, and the radiation energy density is still
 subdominant. That is, in order to end inflation we would have to tune
 the value of the field to the critical value in hybrid inflation. 
 We included the  predictions for the spectral index (dashed lines),
 and the running of  the spectral index (thin solid lines). The
 spectrum is blue-tilted in 
 both cases, and the stronger dissipation is, the closer
 the spectrum is to a scale invariant one.
 For a small coupling $\gamma$ we have that the running is
 positive and goes like $n_S^\prime \simeq 8 (n_S-1)^2/(63 Q_*)$. But
 by increasing the coupling we can reduce the tilt in the spectrum,
 and get a negative running, as shown on the RHS plot. By further
 increasing the coupling we could also revert the tilt of the
 spectrum. However a larger coupling implies a larger value of the
 inflaton vev at horizon crossing, and in the dotted area on the RHS
 plot we have $\phi > m_P$. As with cold inflation, with $\gamma>1$
 inflation is only viable in the weak dissipative regime for values
 $\phi >m_P$. But once in the strong dissipative regime, the extra
 friction helps in keeping the value of the field subplanckian.          
 The scale of inflation can be read from the value of
 $\tilde \gamma^{1/4}=( 3 \gamma  m_P^4/V_0)^{1/4}$ for a given value of $\gamma$. 
And from Eqs. (\ref{qphin}) and (\ref{PRn}), we get the lines  of
constant $Q_*$ in SDWI: 
\be
C_\phi \simeq 5.68 \times 10^{-13} Q_*^{-7/2} \tilde \gamma^2 \,.
\ee
Taking into account the limit $Q_* < 1.3 \times 10^4$ for not having a
too large deviation from gaussianity in the primordial spectrum, we
have:
\be
C_\phi \lesssim 2.27 \times 10^5 \left( \frac{\tilde
  \gamma^{1/4}}{10^4} \right)^8 \,,
\ee
which gives the line ``$Q_{fNL}$'' in Fig. \ref{plotlog}.  
We have also included in the plot the line $Q_* \simeq 1$ dividing the
parameter space for WDWI and SDWI. 

 The predictions for a
 small coupling do not vary as far as $\gamma \lesssim 10^{-2}$, and
 they would be the same than in the LHS plot in Fig. (\ref{plotlog}). 
We can compare these with the ones obtained in supersymmetric cold
hybrid inflation. For example for a value of the coupling 
between the inflaton  and $\chi$ of the order $g \simeq
0.02/\sqrt{N_\chi}$ ($\gamma \simeq 5 \times 10^{-6}$) one gets a scale
of inflation of the order $V_0^{1/4} \simeq 8.5 
\times 10^{14}$ GeV,  and a spectral  index $n_S\simeq 0.98$
\cite{senoguz}. On the other hand in warm inflation, 
for the minimum allowed value of the dissipative parameter $C_\phi
 \simeq 10^6$ we have $\tilde \gamma^{1/4}=(3 \gamma m_P^4/V_0)^{1/4}
 \simeq 200$ which gives  a similar scale of inflation 
 $V_0\simeq 7.5 \times 10^{14}$ GeV for the same value of the
 coupling, but a spectral index $n_S \simeq 1.05$. Such a 
 value for $C_\phi$ can be obtained for example by taking the Yukawa
 coupling $h$ close to its perturbative value, $h \simeq \sqrt{4
   \pi}$, and the multiplicities of order of $N_\chi \simeq N_{decay}
 \simeq 100$. The minimum allowed value for
 $\tilde \gamma$ sets the maximum possible value for the scale of
 inflation in warm inflation. As long as this scale is below $10^{16}$
 GeV, the tensor-to-scalar ratio is suppressed. Even when $\gamma
 \simeq 10^{-2}$ we have $\tilde \gamma \simeq 200$ and therefore
 $V_0^{1/4} \lesssim 10^{16}$ GeV. When $\gamma >1$ by
 demanding $\phi <m_P$ the minimum value for $\tilde \gamma$ increases
 by an order of magnitude, so that again we have $V_0^{1/4} \lesssim 3
 \times 10^{15}$, and therefore no tensor contributions.   

As mentioned before, sugra
corrections coming from  the K\"ahler potential will contribute higher
order terms in the potential, starting with a quartic term with
minimal K\"ahler potential \cite{sugra} $K(\Phi) = |\Phi|^2$:
\be
V= V_0 \left ( 1 + \gamma \ln \frac{\phi}{m_P} + \frac{\phi^4}{8 m_P^4}
+ \cdots \right) \,.
\ee
In CI, the quartic term starts to be non-negligible for a value of the
coupling \cite{senoguz} $g \gtrsim 0.06/\sqrt{N_\chi}$ ($\gamma \simeq
4.6\times 10^{-5}$), such that the
curvature of 
the potential during inflation now becomes positive rendering the
spectrum blue-tilted with 
\be
n_S -1 \simeq 3 \left(\frac{\phi_*}{m_P} \right)^2 - 2 \gamma
\left(\frac{m_P}{\phi_*}\right)^2 \,. 
\ee
However, in warm inflation
the addition of the sugra corrections works in the same direction as
the radiative correction. Taking into account the quartic term
in the slow-roll parameters, the spectral index is given by: 
\bea
n_S -1 &\simeq& \gamma \left(\frac{m_P}{\phi_*}\right)^2 (1 + \delta ) 
 (2 - \gamma  (1 + \delta) ) \,,\;\;\;\; {\rm (WDWI)} \,, \\ 
n_S -1 &\simeq& \frac{3}{14 Q_*} \gamma
\left(\frac{m_P}{\phi_*}\right)^2( 14 - 3 \gamma   - 6( \gamma -1) \delta) \,,\;\;\;\; {\rm (SDWI)} \,, 
\eea
where 
\be
\delta =\frac{1}{2 \gamma}\left(\frac{\phi_*}{m_P}\right)^4 \,.
\ee  
Then, in the small coupling regime $\gamma \ll 1$ sugra corrections
make the coupling more blue-tilted, whereas in the
large coupling regime with $\gamma > 14/3$ it would be more red-tilted,
contrary to what happens in CI. 

\begin{figure}[t]
\centering 
\scalebox{0.4}{\includegraphics{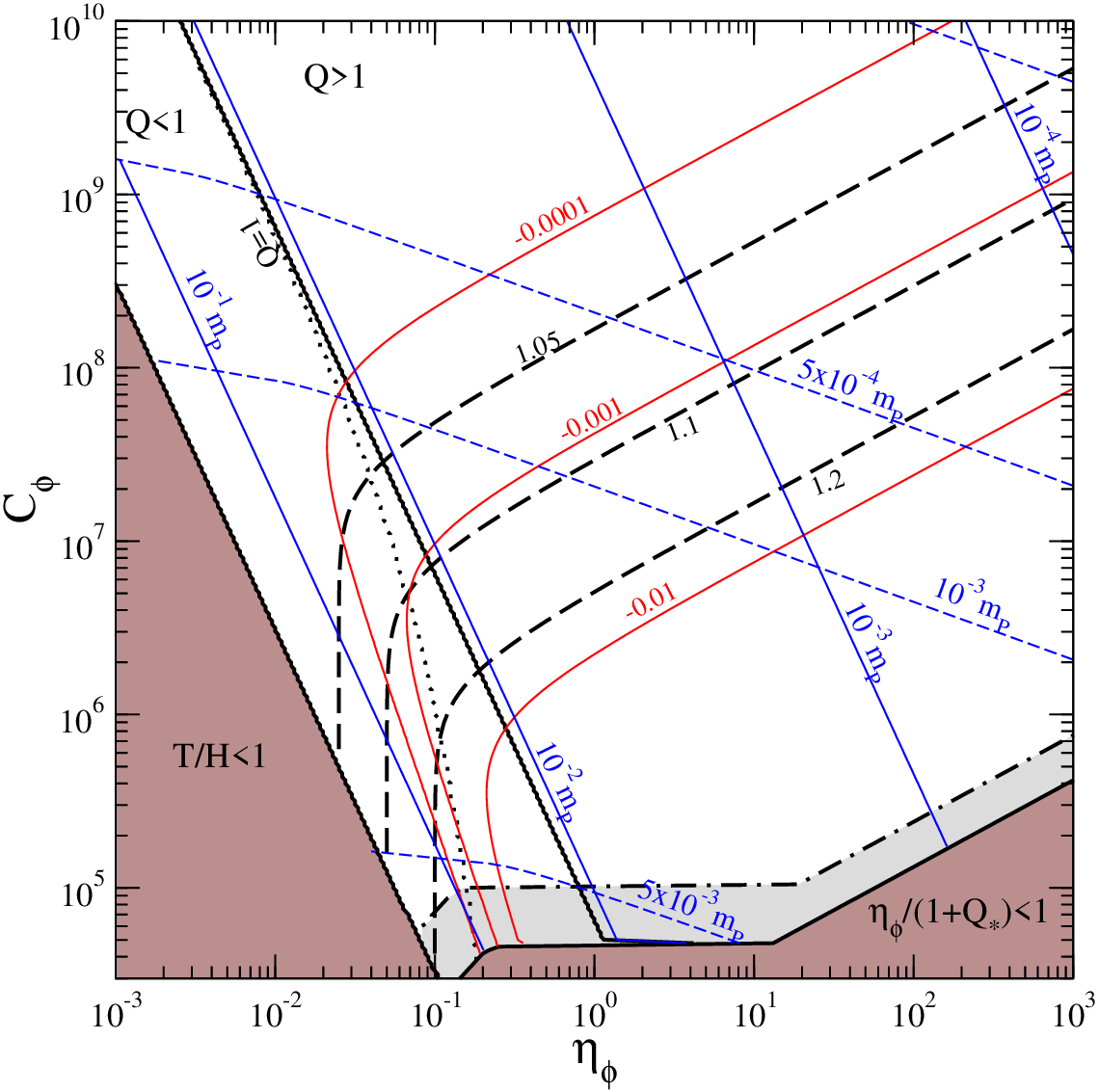}}  
\caption{The $C_\phi$-$\eta_\phi$ plane for the
  quadratic potential, Eq. (\ref{Vn}) with $n=2$ and $\gamma=\eta_\phi$.  
 Same conventions than in Fig. \ref{plotlog}. The thin (blue) lines
 parallel to the line $Q=1$ gives the upper bound on the scale of
 inflation $V_0^{1/4}$ demanding that $\rho_R < V_0$. The thin (blue)
 dashed lines are the upper limit from the constraint $\phi < m_P$. }
\label{plotqd}
\end{figure}

For the quadratic potential, Eq. (\ref{Vn}) with $n=2$, the allowed
parameter space in the plane $C_\phi$-$\gamma$ ($\gamma=\eta_\phi$) compatible 
with the primordial spectrum is given in Fig. (\ref{plotqd}). Because
now $T/H$ and $\phi/T$ 
decrease during inflation, when varying $C_\phi$ and $\gamma$ we have to check 
that we can get at least 40 (60) e-folds before $T/H <1$ or $\phi/T < 10$. 
Indeed we always first reach $T=H$, and this corresponds to the
shaded area on the LHS corner of the plot. In this case the low-$T$
requirement $m_\chi/T \gtrsim 10$ is satisfied in general for $g
\gtrsim 2 \times 10^{-3}$ ($g \gtrsim 10^{-4}$ for $C_\phi > 10^5$). 
On the other hand, when
$\gamma \gg 1$ the constraint comes from having a small enough slow-roll
parameter to start with at horizon crossing (the bottom RHS corner).   
The lower limit on the dissipative parameter for having at least 40
e-folds of WDWI (SDWI) is $C_\phi \simeq 3\times 10^4$ ($5 \times 10^4$),
which can be easily obtained with multiplicities of the order of 10-100. 
Given that $T$ decreases, the radiation energy density never dominates
during inflation, and in order to end inflation after the 40 (60)
e-folds we would have to tune the value of the inflation field to the
critical value. The spectral 
index (dashed lines) is always larger than 1, going like $n_S \simeq 1
+ 2 \gamma$ in  WDWI (same than in cold inflation), but like $n_S
\simeq 1 + 15 \gamma/(7Q_*)$ in SDWI, i.e., it decreases by
increasing $C_\phi$ 
for constant $\gamma$. The running of the spectral index is always
negative (thin solid lines), in both regimes. In WDWI it is the contribution
proportional to $\gamma^2 Q_*$ the dominant one (see Table
\ref{table4}). In SDWI we have at leading order $n^\prime_S \simeq
-2 (n_S-1)^2/15$. For example for $C_\phi \simeq 10^6$ and $\gamma
\simeq 1$ we can have a large negative running $n^\prime_S \simeq
-0.01$, but with a rather large spectral index $n_S \simeq 1.27$,
which seems excluded by the latest data. By increasing the dissipative
parameter an order of magnitude, which could be accomplished by
doubling the multiplicities, we can reach $n_S\simeq 1.1$ for $\gamma
\simeq 1$, although with $n_S^\prime \simeq -0.001$. The spectral
index is still too blue-tilted, but within the allowed range by WMAP
data \cite{WMAP5}.   

The predictions related to the  spectrum are
rather insensitive to the value of $V_0$. They depend on $\gamma$ and
$Q_*$, and the latter in turn depends on $\gamma$ and $C_\phi$. In
particular, the lines of constant $Q_*$ respectively in WDWI and
SDWI are given by: 
\bea
C_\phi &\simeq& 1.4 \times 10^4 Q_*^{1/3} \gamma^{-2} \,,\;\;\; ({\rm 
  WDWI}) \,,\\
C_\phi &\simeq& 1.4 \times 10^4 Q_*^{5/2} \gamma^{-2} \,,\;\;\; ({\rm
  SDWI}) \,.
\eea
By varying the scale of inflation we just rescale the ratio $\rho_R/V_0$
at the start of inflation,
the initial value of the field $\phi/m_P$, and the prediction
for the ratio $r$. Demanding that $\rho_R$ does not dominate at least
at horizon crossing  gives us the maximum allowed value for each pair
 of values ($C_\phi, \gamma)$), which in
Fig. \ref{plotqd} is given by the thin (blue) lines parallel to the
$Q=1$ line. Well in WDWI, and SDWI, we have respectively:
\bea
V_0^{1/4} &\simeq& 1.26  Q_*^{-1/3} m_P \,,\;\;\; ({\rm 
  WDWI}) \,,\\
V_0^{1/4} &\simeq& 1.26 \times 10^{-2} Q_*^{-1/2} m_P\,,\;\;\; ({\rm
  SDWI}) \,.
\eea
Then, the limit from non-gaussianity is approximately given in this case by
the line $V_0^{1/4} \simeq 10^{-4} m_P$. 

On the other hand, the dashed (blue) thin lines are the
maximum value of the potential for having $\phi_* \leq m_P$, 
with $V_0^{1/4} \lesssim 0.16 m_P (C_\phi^3 \gamma)^{-1/10}$. 
By using the minimum $C_\phi$ value and $\gamma$ read from the plot, the latter
condition sets the upper bound for the potential at $V_0^{1/4} \simeq
1.2\times 10^{16}$, which gives an upper bound on the tensor-to-scalar
ratio, Eq. (\ref{rVn}), $r \lesssim 0.017$. With respect to the
inflaton mass, for example in WDWI
with $C_\phi \simeq 10^6$ and $\gamma \simeq 0.05$  ($n_S \simeq 1.1$)
we have the upper bound $V_0 \lesssim 7.2 \times 10^{15}$ GeV, which
gives the upper bound on the inflaton mass $m_\phi \lesssim 1.2 \times
10^{12}$ GeV. On the other hand, in CI for the same value of $\gamma$,
from the amplitude of the spectrum the scale of the potential 
is given by $V_0^{1/4} \simeq 4.8 \times 10^{14} g^{-1/2}$ GeV, and
therefore $m_\phi \simeq 2.15\times 10^{10} g^{-1}$ GeV, where $g$ is
the coupling between the inflaton and the $X$ field. Having in CI similar
values to those of WDWI for $V_0$ and $m_\phi$  requires a small value of
the coupling $g \simeq 4 \times 10^{-3}$, whereas in warm inflation the
predictions do not depend on this coupling as far as $g \gtrsim
2\times 10^{-3}$.

\begin{figure}[t]
\centering 
\begin{tabular}{cc}
\scalebox{0.4}{\includegraphics{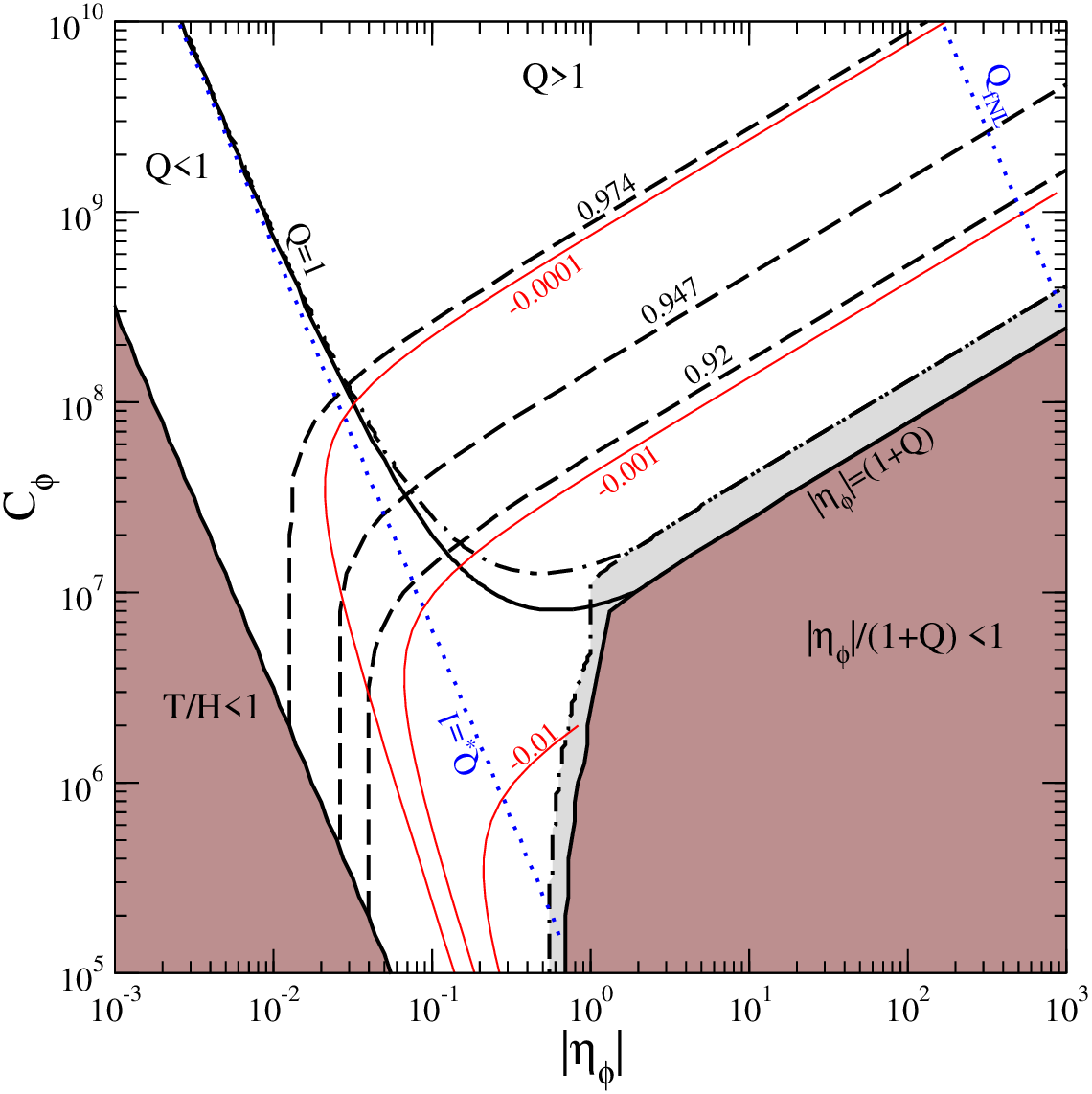}}  &
\scalebox{0.4}{\includegraphics{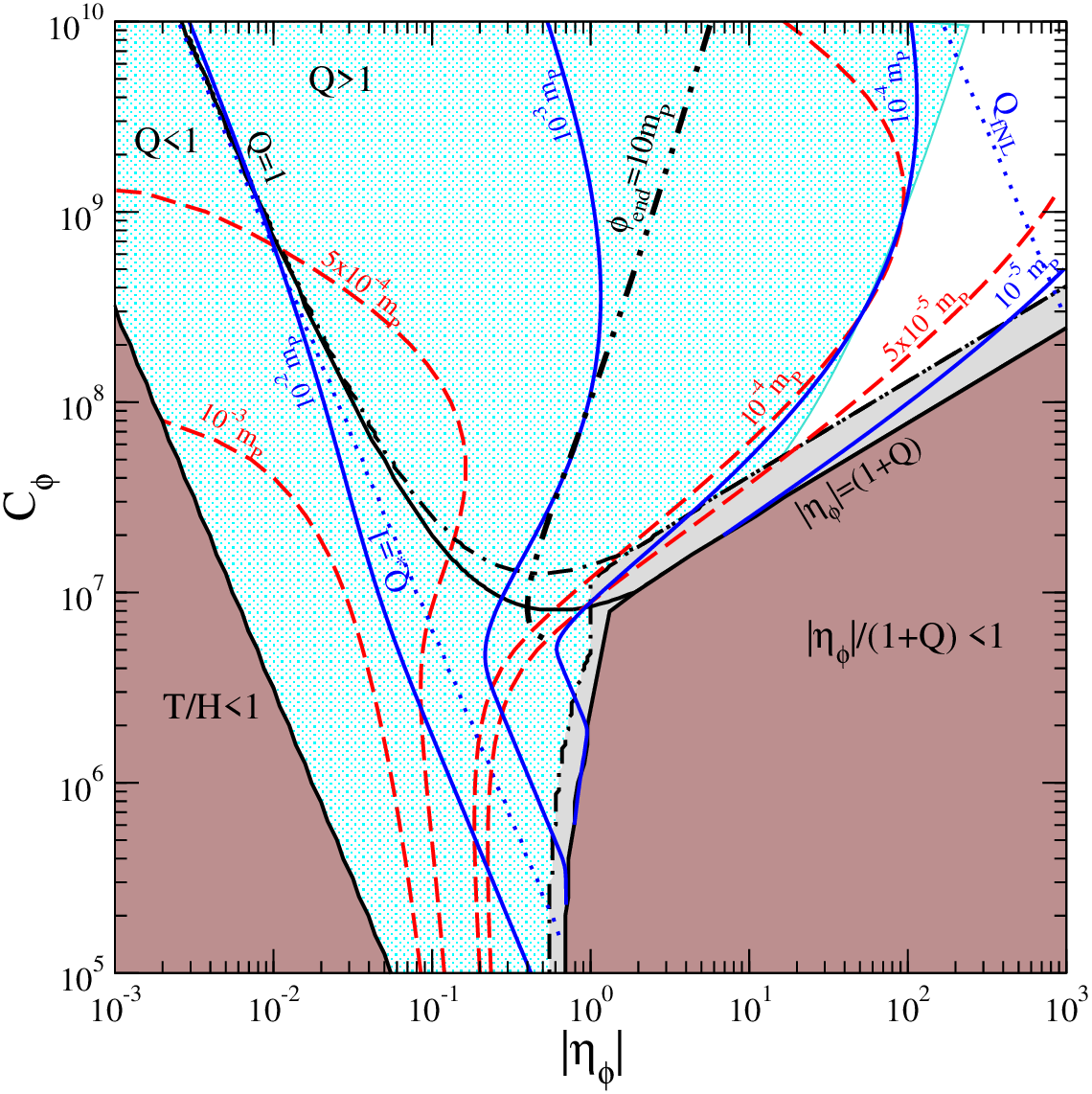}}  
\end{tabular}
\caption{The $C_\phi$-$|\eta_\phi|$ plane for the
  hilltop potential, Eq. (\ref{Vn}) with $n=2$ and negative squared
  mass $\gamma=\eta_\phi <0$.  Same conventions than in Fig. \ref{plotlog}. The
  curved thick solid (dot-dashed) line labelled $Q=1$ to the left and
  by $|\eta_\phi |=(1+Q)$  to the right separates
  WDWI from SDWI when
  $N_e=40$ (60).  The (blue) dotted line gives the value of $Q_*$ at
  horizon crossing, with that labelled $Q_{fNL}$ being the upper limit from
  non-gaussianity. On the LHS plot we have included the lines of
  constant $n_S$ (dashed-lines) and $n^\prime_S$ (thin red solid
  lines). On the RHS plot the (blue) solid lines gives the the upper
  bound on the scale of inflation $V_0^{1/4}$ demanding that $\rho_R < V_0$ at
  least during 40 e-folds of inflation, while the (red) dashed lines are 
  the corresponding value for having $\phi = m_P$ at the end of
  inflation. By setting $\rho_R=V_0$,  in the dotted shaded area  we 
  have $\phi_{end} > m_P$, while to the left of the thick
  dot-dot-dashed line we have   $\phi_{end} >  10m_P$. }    
\label{plotht}
\end{figure}

\section{Hilltop models}
\label{wihilltop}

Hilltop models, or new inflation models, are given by the potential
during inflation \cite{Linde:1981mu,hilltop}:
\be
V=V_0 \left( 1 - \frac{|\gamma|}{2} \left(\frac{\phi}{m_P}\right)^2
\right) + \cdots\,, \label{Vht}
\ee
where the dots account for higher order terms. This is no more than
a quadratic potential of the kind considered in the previous section,
but with a negative squared inflaton mass. In this model the inflaton
starts at small values close to zero, and rolls off the top of the
potential hill during inflation. When the field reaches a large enough
value, higher order terms in the potential start contributing and
inflation ends soon after. The mechanism for ending inflation differs
from the quadratic hybrid model studied in the previous section, but
for example the predictions for the spectral index, running and
tensor-to-scalar ratio are the same replacing now $\gamma$ with
$-|\gamma|$. Thus, the primordial spectrum is red-tilted in both CI and
WI, with a negative running (see Table \ref{table4}). But given that
the field now increases as inflation proceeds, during WI the
dissipative ratio $Q$ decreases instead, while $T$ and the ratio
$\phi/T$ keep increasing. In this case we can have either the 40(60)
e-folds of inflation in WDWI or SDWI, or a transition from
SDWI towards WDWI before the end of inflation. 
And once the system enters in the low-$T$ regime, it remains there
until the end of inflation.

Warm hilltop inflation has been studied in detail in
Ref. \cite{BuenoSanchez:2008nc}, and in Fig. \ref{plotht} we
summarised the results. The analyses is done similarly to that of the
potentials in the previous section. We show the values in the plane
$C_\phi$- $|\gamma|$ consistent with the amplitude of the primordial
spectrum, which fix the initial conditions for warm inflation. The LHS
shaded corner is excluded because  $T \leq H$ at horizon
crossing. 
The condition $m_\chi/T \gtrsim 10$ then only requires  
 a coupling $g \gtrsim  10^{-4}$ for $C_\phi \gtrsim 10^5$.   
The RHS dark (light grey) shaded area is excluded because of
the violation of the slow-roll condition before reaching $N_e=40$ (60)
e-folds. The lines of constant $n_S$ and $n^\prime_S$ are given on the
LHS plot.  The solid line labelled $Q=1$ divides the plane for WDWI (to
the left) and SDWI (to the right). Therefore, in order to have at
least 40 e-folds in SDWI it is required $C_\phi \gtrsim
10^{7}$. Nevertheless, the spectral index for such a value would be
too red-tilted, but for example with $C_\phi \sim 10^8$ and $\gamma
\sim 1$ we can have $n_S$ within the 1$\sigma$ range given in
\cite{WMAP5}, $n_S = 0.96 \pm 0.013$.

As in the previous section, those
predictions are independent of the scale of inflation, but by
increasing the value of the height of the potential the radiation
energy density also increases. Therefore, on the RHS plot the solid
lines gives the upper limit on the scale of inflation such that the
radiation does not dominate during inflation. For each point in the
allowed plane, by setting $V_0$ to its upper bound we will end
inflation after 40 e-folds, going smoothly from a vacuum dominated
universe into a radiation dominated one. The line $V_0 = 10^{-2}
m_P$  set the limit for not having a too large tensor-to-scalar
contribution $r \lesssim 0.22$. 
We can also demand that inflation ends
before the inflaton field reaches the Planck mass, and this gives the
upper limit on $V_0^{1/4}$ shown by the dashed lines. The latter constraint
is more restrictive than the one on the radiation energy density. 
Indeed, in the dotted shaded area on the RHS plot by keeping the value
of $V_0$ such that $\rho_R = V_0$ after 40 e-folds we have that
$\phi_{end} > m_P$.  If we relax the condition on the
inflaton vev, to the right of the dot-dot-dashed line the value of the
inflaton field is still below $10 m_P$. 

\section{Warm inflation and model building}
\label{wimodelb}

We have studied the parameter space that leads to a period of warm
inflation in the low-$T$ regime, for three kinds of generic
single-field inflationary 
potentials. Implicitly, we are dealing with supersymmetric models for
which the interactions relevant for the dissipative dynamics are given
in Eq. (\ref{superpot}), and the inflaton potential can be derived
from Eq. (\ref{WPhi}). Supersymmetry ensures that quantum and thermal
corrections to the effective potential are under control
\cite{br05,Hall:2004zr}. On the other 
hand, the inflaton potential can be understood from the point of
view of an effective field theory, with the inflationary potential
dominated by the first few terms of a series, such that higher order
terms are suppressed by the increasing inverse power of the cutoff
scale. Without a specific setup to fix the cutoff, this naturally
will be given by the Planck scale $m_P$. We have been
considering models of inflation dominated by the first term of
the expansion, but higher order terms appear naturally in supersymmetric
models either by taking into account non-renormalizable interactions
in the superpotential, and/or sugra corrections from the K\"ahler
potential. The interactions in the superpotential can always be
restricted by the use of symmetries, local, global or discrete. 
The pattern of interactions in Eq. (\ref{superpot}) can follow for
example by considering $X$ and $Y$ to be charged under some GUT group, 
but taking $\Phi$ to be a gauge singlet. We can further impose a
discrete symmetry to select the leading term in the inflaton
superpotential. During inflation the breaking of the discrete symmetry
can lead to the formation of topological defects like domain walls
\cite{walls}, but these will be inflated away. For the standard
superpotential for susy hybrid 
inflation, it is customary to impose a $R$-parity symmetry on the
superpotential \cite{shafi} in order to eliminate higher order terms for the
inflaton apart from the linear term. But in general one can always
expect non-renormalizable terms at some higher order in the potential,
with generic couplings of the order of one. Such terms can be safely
ignored in models where $\phi_{end} < m_P$, i.e., small field models. 

Chaotic models do not fall into that category, and cold inflation is
realised only for field values larger than $m_P$, with $\phi_* \simeq
\sqrt{2 N_e n m_P}$ and $\phi_{end} \sim m_P$. Once we have   
selected by the appropriate symmetry the power in the superpotential,
we have to deal with sugra corrections from the K\"ahler
potential. Just the simplest example, with a quadratic potential $W=M \Phi^2$ a
minimal K\"ahler potential $K(\Phi) = |\Phi|^2$ induces higher order
terms  with 
\be
V \simeq M^2 |\Phi|^2 ( 1 + \frac{|\Phi|^6}{m_P^6} + \cdots ) \,
\label{sugraqd}
\ee          
which invalidate slow-roll inflation. The choice of the K\"ahler
potential compatible with inflation  can be dictated again by imposing
some symmetry \cite{Goncharov:1983mw}, either for example a shift
symmetry like in 
Refs. \cite{shift}, or a Heisenberg symmetry used to ensure a flat
enough potential \cite{Gaillard:1995az}. In
addition, the use of the Heisenberg symmetry implies the introduction
of at least an extra degree of freedom, the modulus field. The
presence of such fields, the moduli, is common    
in the low energy effective 4-dimensional sugra model derived from the
compactification of the higher dimensional string theory. 
Although a suitable chaotic inflaton potential can be achieved by a
specific choice of the K\"ahler potential \cite{Murayama:1993xu},
moduli fields need to be stabilised  and fixed before hand in order to allow 
slow-roll inflation \cite{Ellis:2006ara}, which could be done at
the expenses of some degree of fine-tuning in the potential.
But see  for example Ref. \cite{stringchaotic} for chaotic models
derived in string theory with the higher order corrections under
control.  

Large field models, or chaotic models, are interesting because of the
non-negligible prediction for the tensor-to-scalar ratio, $r \gtrsim
0.01$, within the range that could be detected by the current and
future CMB experiments like Planck \cite{planck} and CMBPol
\cite{cmbpol}. 
And indeed, already
the current limit on $r$ disfavours quartic chaotic potentials and
higher powers. On the other hand, in chaotic WI one predicts a smaller
value for $r$, due to the suppression by $T/H$ and the dissipative
ratio $Q$, Eq. (\ref{rwi}). Therefore, even only having WDWI, with 
$Q < 1$ at the time of horizon crossing, the suppression due to the
temperature is enough to render the quartic (and sixth order)
potential compatible with observations. This would happen for values
of the dissipative parameter $C_\phi \simeq O(10^6)$, which as
mentioned previously only requires multiplicities of the order of
$O(100)$ or so for the fields. Still inflation takes place with values
of the field $\phi > m_P$, and like in the CI scenario we have to deal  with the
problem of higher order corrections. By increasing the dissipative
parameter by a couple of orders of magnitude, we are in SDWI with
$Q\gg 1$, and the extra friction slows down enough the motion of the field
for having inflation with a subplanckian inflaton vev. In this
case higher order corrections can be easily kept under
control\footnote{One would need to check how many terms in the
  potential expansion are relevant at the time of horizon crossing,
  and their impact on the inflationary predictions, 
  but still they do not spoil the slow-roll conditions.}, and even the
simplest model Eq. (\ref{sugraqd}) gives rise to a viable model.      
Inflation ends when the slow-roll conditions Eq. (\ref{wslowroll}) do
not longer hold, similarly to cold inflation. But in this regime, the
extra suppression due to the dissipative ratio gives a negligible
prediction for the tensor-to-scalar ratio, as it would be expected
from small-field models (i.e., subplanckian field models).

Hybrid models of the kind studied in Sect. \ref{sect2} are small field models,
with $\phi < m_P$ during inflation, and therefore one expects by
definition to have a better control on the potential expansion. Nevertheless,
when embedding the susy potential into the sugra framework, due to the
constant term dominating the potential energy, sugra corrections
generate a mass term for all scalars of the order of the Hubble
parameter, and in particular for the inflaton field, the so-called
``eta'' problem \cite{Dine}: 
\be
V_{sugra}\simeq e^{K(\Phi)}( V_0 + \cdots ) \simeq V_0 +
\frac{V_0}{m_P^2} K_{\Phi \Phi*} |\Phi|^2 + \cdots \sim V_0 + c H^2
|\Phi|^2 + \cdots \,,
\ee
where $K(\Phi)$ is the K\"ahler potential, and $K_{\Phi\Phi*}$ the
second derivative with respect to the inflaton field. Without tuning
any parameter, one expects $|c|\simeq O(1)$, and then $|\eta_\phi| \sim 
O(1)$. Again, by taking into account some particular choice of
superpotential + K\"ahler potential, like the standard susy hybrid
inflation with minimal K\"ahler, different terms conspire to cancel
out the quadratic term, although quartic and higher order terms will
be present nonetheless. In this model the flat potential at
tree-level is lifted by the 1-loop corrections giving rise to a log
potential. However, quartic corrections become important for
values of the coupling of the order of $g \simeq 0.06 /\sqrt{N_\chi}$,
rendering the spectrum too blue-tilted for larger values, and making
$\phi_* > m_P$, so that we loose control on the corrections. Going
beyond the minimal choice for $K$, the $\eta$ 
problem is reintroduced, which requires as usual a mild tuning of the
parameters to achieve inflation \cite{BasteroGil:2006cm}; and working with
effective sugra models coming from string compactifications the $\eta$
problem is directly related to the moduli stabilisation problem  
\cite{Brax:2006ay}. Due to the coupling between modulus and inflaton
through non-canonical kinetic terms, a small variation of the moduli field
gives rise to a too large contribution to the $\eta$ parameter, and a
proper choice of the K\"ahler potential is mandatory in order to have
inflation.     

In a setup where we have dealt with moduli stabilisation, dissipation
will help with the remaining $\eta$ problem and higher order sugra
corrections. When the inflationary dynamics is dominated by the log
term, we can take larger
values of the coupling $g > 0.06/\sqrt{N_\chi}$ ($\gamma > 4.6 \times
10^{-5}$) and still the quartic sugra corrections will be subdominant,
with $\phi$ well below $m_P$.  The value of the coupling at which the
inflaton approaches $m_P$ is larger, but still by decreasing
accordingly the height of the potential, i.e., by increasing $\tilde
\gamma= 3 \gamma m_P^4/V_0$ we can keep the field within the range
of the effective field description (see the RHS plot in
Fig. \ref{plotlog}). The main 
point is that we have more freedom to choose the value of the scale of
inflation, not being directly fixed by the amplitude of the primordial
spectrum due to its different thermal origin. In this case the
constraint from $P_{\cal R}$ fixes the values at horizon crossing for the
dissipative ratio $Q_*$ and therefore the value of the inflaton
field. Nevertheless, warm susy hybrid inflation predicts
a blue-tilted spectrum, although the curvature of the potential is
negative, in both regimes, weak and strong dissipation, except when
the coupling is large enough  such that the contributions from
$\epsilon_\phi$ are non negligible (see Table \ref{table1}).  
On the other hand, when sugra corrections generate a positive squared
mass term for the inflaton field, we can have the scenario given in
Fig. (\ref{plotqd}). Without the need of suppressing the mass,
inflation can proceed in SDWI with $\gamma = \eta_\phi = m^2_\phi m^2_P/V_0
>1$. The spectrum is blue-tilted, but a large enough dissipative
parameter can render it closer to a scale-invariant one. In this two
hybrid examples, inflation ends when reaching the critical value
$\phi_c$. We will have to tune $\phi$ to such a value $N_e\sim 40-60$ 
e-folds after horizon crossing. But given the relative freedom we have now in
choosing the height of the potential this does not impose a severe
constraint to the model. Indirectly this is the same as in cold
inflation, the amplitude of the spectrum fixes one of the parameters
of the model, which in our case ends up being the critical value of the
field. Also the same as in CI, the tensor-to-scalar contribution is
negligible, except in some corner of the parameter space for the log
potential with the minimum allowed value for $\tilde \gamma$, for which
we could get $r \simeq 0.01$.
   
When the quadratic sugra term gives rise to a negative squared mass
term as the dominant term \cite{izawa}, we may have hilltop inflation, with
the field starting near the top of the potential and evolving towards
larger values. The problem in this case is how to end inflation
without running into too large values of the inflaton field, i.e., by
keeping $\phi < m_P$, which implies that the potential has to steepen
sharply after $N_e$ e-folds (see for example
\cite{hilltop2}). Nevertheless, CI hilltop models with the quadratic
term seem to be disfavoured already by observations \cite{alabidi},
and with a positive power it is required $n \gtrsim 3$. This again
leaves open the question of how to suppressed the lower terms when 
$\phi \ll m_P$ \cite{sarkar}. For WI, we have identified the parameter
space that leads to enough inflation consistent with observations (see
Fig. (\ref{plotht})), and even with a mass parameter $|\gamma| = |\eta_\phi|
\simeq 1 $ of the order expected in sugra theories, inflation in
SDWI is a viable option. Although the problem still remains of how to end
inflation with $\phi < m_P$. A nice solution would be to
consider the model as an inverted hybrid model
\cite{invertedhyb,invertedhyb2}, with the field rolling now from small
values towards the critical value, and like in the standard hybrid model tune
the value of $\phi_{end} = \phi_c$, with a low enough scale of
inflation, below the GUT scale $V_0^{1/4} \lesssim 10^{16}$ GeV. For
example the vacuum energy can originate from the $F$-term leading to
susy breaking, with $V_0 \simeq m_{3/2}^2 m_P^2 \simeq (10^{11} \, {\rm
GeV})^4$, with $m_{3/2}\sim 1$ TeV being the gravitino mass. On the
other hand, during WI the radiation energy density 
increases, and inflation can end when $\rho_R \simeq V_0$, with a
smooth transition from vacuum domination to a radiation dominated
universe. The corresponding values of the vacuum energy required are
given on the RHS plot in Fig. (\ref{plotht}). However, typically we
will have $\phi_{end} > m_P$ (dotted shaded area). In CI this kind of
scenario could be implemented in the context of natural inflation
\cite{natural}, where the inflaton field is a pseudo Nambu-Goldstone
boson (PNGB), with a periodic potential and a scale of spontaneous
breaking $f > m_P$. The leading term in the expansion of the periodic
potential leads to the negative quadratic term. In the context of warm
inflation it would remain to be seen if such a field could couple to
other fields in the model with the pattern of interactions required for the
dissipative dynamics.

\section{Summary and Future Challenges}
\label{summary}

To summarize, we have revised the warm inflation dynamics,  with the
dissipative coefficient in the low-$T$ regime derived in the
close-to-equilibrium aproximation \cite{BGR} as computed in
\cite{mx,Berera:2008ar}. Dissipation 
of the inflaton energy into light degrees of freedom follows from 
the two-stage mechanism in Ref. \cite{BR1,br,br05}, with the
interactions given in Eq. (\ref{superpot}). For the inflationary
potential we have focused on models where the primordial spectrum
originates from the inflaton quantum fluctuations and the thermal
fluctuations of the energy density dissipated. As particular models, we
have studied supersymmetric models where the relevant inflationary
potential is dominated by the first term in the expansion of the effective
potential, but the analyses presented in Sect. \ref{sect2} is rather general
and it can be easily extended to other models. Chaotic models up
to a power $n<14$, and hybrid-like models with a logarithmic or a
quadratic potential, always lead to a period of strong dissipation for a
large enough dissipative coefficient. This facilitates the embedding of
the models into a sugra framework as explained in Sect. \ref{wimodelb}. 
Apart from providing a solution to the standard $\eta$ problem, this
would open for example the  possibility of having sugra inflation without
fully stabilising the moduli fields, but just keeping them in a
slow-roll trajectory. The large $\eta_\phi$ induced by the motion of
the moduli can be counteracted by the dissipative dynamics.   

Remarkable for chaotic models, even
with weak dissipation, the different thermal origin of the primordial
spectrum when compared to the cold inflation predictions render the
tensor-to-scalar ratio consistent with observations for quartic
models. On the contrary, hybrid models tend to give rise to a
blue-tilted spectrum even when the curvature of the potential is
negative. Reverting the sign of the quadratic term in hybrid models we
revert the evolution of the inflation field from small to large
values, the hilltop model,  and the behavior of the dissipative parameter,
with the tilt of the spectrum now being again red-tilted. Warm hilltop
models when only considering one dominating term in the potential may
end inflation with a too large value of the inflaton field. On the
other hand, given the freedom we have now to choose the scale of
inflation, not being directly dictated by the amplitude of the
primordial spectrum, as an alternative this kind of models could be
embedded without difficulty into an inverted hybrid model, where
inflation ends when reaching the critical value  as in
the standard hybrid case.    

A feature generally found of warm inflation models so far has been
the requirement of many fields, usually in the hundreds or more.
This is in contrast to cold inflation models which usually
involve just a few fields.  In the high energy regime, most particle
physics models exhibit a large proliferation of fields,
with the ultimate example being that of string theory, where field
numbers can become huge.  Thus warm inflation models possibly have
a natural setting at high energy, based on what model building
has found out about field content of
particle physics models at high energy scales.
Some studies have been made of warm inflation models
in the context of string theory \cite{bk}
(a variation of the warm inflation picture in string models also has been
explored \cite{battefeld}), and further analysis may prove interesting.

The results presented here assume that the relevant $N_e$
e-folds of inflation take place in the low-$T$ regime, such that the
equations of motion can be solved analytically. In the intermediate or
high $T$ regime, with $m_\chi \gtrsim T$, the dependence of the
dissipative coefficient $\Upsilon$ changes, for example with $\Upsilon
\propto T \ln T$ ($m_\chi \ll T$), or $\Upsilon \propto T^{-1} \ln T$
($m_\chi \ll h T$) \cite{BGR,mx,Berera:2008ar}, 
but still with this $T$ dependence
we could have some extra e-folds of inflation, and taking into account
this regime can enhance the parameter space for warm inflation. 
For example for monomials potentials this would happen at the end of
inflation, and having a smaller no. of e-folds in the low-$T$ regime
means that horizon crossing may take place in SDWI for smaller
values of the dissipative parameter $C_\phi$, which implies a smaller
prediction for the tensor-to-scalar ratio. On the contrary, for
hilltop models as those in Sect. \ref{wihilltop}, the
intermediate/high $T$ regime will occur at the beginning of
inflation. This opens up the possibility of having horizon
crossing in that regime, and different predictions for the spectral
index, running, and tensor-to-scalar ratio than those given in
Sect. \ref{wihybrid}. 

On the other hand going beyond the low-$T$ approximation one would have to
take into account the thermal corrections to the inflaton effective
potential, which we have neglected so far. With the two-stage 
mechanism for dissipation, those are due to the contribution of the 
self energies of the heavy fields $\chi$ and $\psi_\chi$, which in
turn receive thermal corrections due to the interaction with the light
fields (the thermal bath) $y$ and $\psi_y$ \cite{Hall:2004zr}. In a
hybrid model where inflation ends by a phase transition once
$m_{\chi_R}^2<0$, the thermal corrections will tend to keep this
squared mass positive and delay the phase transition. All effects
considered, they tend to increase the duration of inflation, and it
remains to be seen how this will affect the time of horizon crossing
and the primordial spectrum.

\section*{Acknowledgements} 
The work of M.B.G. is partially supported by the M.E.C. under contract
FIS 2007-63364 and by the ``Junta de Andaluc\'{\i}a'' group FQM 101. 
A.B. is supported by STFC.

\end{document}